\documentclass[twocolumn]{aastex631}

\usepackage[shortlabels]{enumitem}

\usepackage[normalem]{ulem}
\usepackage{ragged2e}

\shorttitle{YSOs variability with Rubin LSST}
\shortauthors{Bonito and Venuti al.}
\graphicspath{{./}{figures/}}

\begin{document}

\title{Young stellar objects, accretion disks, and their variability with Rubin Observatory LSST}

\author[0000-0001-9297-7748]{R. Bonito}\thanks{These authors contributed equally to this work.}
\affiliation{INAF -- Osservatorio Astronomico di Palermo, P.zza del Parlamento 1, 90134 Palermo, Italy}

\author[0000-0002-4115-0318]{L. Venuti}\thanks{These authors contributed equally to this work.}
\affiliation{SETI Institute, 339 Bernardo Avenue, Suite 200, Mountain View, CA 94043, USA}
\affiliation{CSSM and Department of Physics, University of Adelaide, North Terrace campus, Adelaide, SA 5005, Australia}

\author[0000-0003-4596-2628]{S. Ustamujic}
\affil{INAF -- Osservatorio Astronomico di Palermo, P.zza del Parlamento 1, 90134 Palermo, Italy}

\author[0000-0003-2874-6464]{P. Yoachim}
\affil{Department of Astronomy, University of Washington, 3910 15th Ave. NE, Seattle, WA 98195, USA}

\author[0000-0001-6279-0552]{R.~A. Street}
\affil{Las Cumbres Observatory, 6740 Cortona Drive, Suite 102, Goleta, CA 93117, USA}

\author[0000-0002-8893-2210]{L. Prisinzano}
\affil{INAF -- Osservatorio Astronomico di Palermo, P.zza del Parlamento 1, 90134 Palermo, Italy}

\author[0000-0002-5380-549X]{P. Hartigan}
\affil{Physics and Astronomy Dept., Rice University, 6100 S. Main, Houston, TX 77005, USA}

\author[0000-0002-3010-2310]{M.~G. Guarcello}
\affil{INAF -- Osservatorio Astronomico di Palermo, P.zza del Parlamento 1, 90134 Palermo, Italy}

\author[0000-0002-3481-9052]{K.~G. Stassun}
\affil{Department of Physics \& Astronomy, Vanderbilt University, 6301 Stevenson Center Lane, Nashville, TN 37235, USA}

\author[0000-0002-7035-8513]{T. Giannini}
\affil{INAF -- Osservatorio Astronomico di Roma, via Frascati 33, 00078 Monte Porzio Catone, Italy}

\author[0000-0002-5077-6734]{E.~D. Feigelson}
\affil{Department of Astronomy \& Astrophysics, Pennsylvania State University, 525 Davey Laboratory, University Park, PA 16802, USA}

\author[0000-0001-8876-6614]{A. Caratti o Garatti}
\affil{INAF -- Osservatorio Astronomico di Capodimonte, via Moiariello 16, I-80131 Napoli, Italy}
\affil{Dublin Institute for Advanced Studies, School of Cosmic Physics, Astronomy \& Astrophysics Section, 31 Fitzwilliam Place, Dublin 2, Ireland}

\author[0000-0003-2836-540X]{S. Orlando}
\affil{INAF -- Osservatorio Astronomico di Palermo, P.zza del Parlamento 1, 90134 Palermo, Italy}

\author[0000-0002-2577-8885]{W.~I. Clarkson}
\affil{Department of Natural Sciences, University of Michigan-Dearborn, 4901 Evergreen Road, Dearborn, MI 48128, USA}

\author[0000-0003-0948-6716]{P. McGehee}
\affil{SLAC National Accelerator Laboratory, 2575 Sand Hill Rd, Menlo Park, CA 94025 USA}

\author[0000-0001-8018-5348]{E.~C. Bellm}
\affil{DiRAC Institute, Department of Astronomy, University of Washington, 3910 15th Avenue NE, Seattle, WA 98195, USA}

\author[0000-0002-8916-1972]{J.~E. Gizis}
\affil{Department of Physics and Astronomy, University of Delaware, 217 Sharp Lab, Newark, DE 19716, USA}

\begin{abstract}

Vera C. Rubin Observatory, through the Legacy Survey of Space and Time (LSST), will allow us to derive a panchromatic view of variability in young stellar objects (YSOs) across all relevant timescales. Indeed, both short-term variability (on timescales of hours to days) and long-term variability (months to years), predominantly driven by the dynamics of accretion processes in disk-hosting YSOs, can be explored by taking advantage of the multi-band filters option available in Rubin LSST, in particular the $u,g,r,i$ filters that enable us to discriminate between photospheric stellar properties and accretion signatures. The homogeneity and depth of sky coverage that will be achieved with LSST will provide us with a unique opportunity to characterize the time evolution of disk accretion as a function of age and varying environmental conditions (e.g. field crowdedness, massive neighbors, metallicity), by targeting different star-forming regions.
In this contribution to the Rubin LSST Survey Strategy Focus  Issue, we discuss how implementing a dense observing cadence to explore short-term variability in YSOs represents a key complementary effort to the Wide-Fast-Deep observing mode that will be used to survey the sky over the full duration of the main survey ($\approx$10~years). The combination of these two modes will be vital to investigate the connection between the inner disk dynamics and longer-term eruptive variability behaviors, such as those observed on EXor objects.
\end{abstract}


\section{Introduction} \label{sec:intro}

Vera C. Rubin Observatory, through the Legacy Survey of Space and Time (hereafter Rubin LSST), will be the ideal facility to allow the investigation of variability in young stellar objects (YSOs), systems composed by a young ($\lesssim 5-10$~Myr) star, its surrounding disk, jets, and envelopes. Crucially, the duration of the LSST survey will encompass all the very different timescales relevant to YSO variability, which bear the imprints of diverse phenomena taking place within the star-inner disk environment: mass accretion from the inner disk onto the star, magnetic activity including energetic manifestations like stellar flares and coronal mass ejections, and geometric effects such as rotational luminosity modulation by surface spots or extinction events due to inner disk warps along the line of sight to the source. These processes exhibit variations on different timescales from short-term (hours\footnote{See also \citet{Bellm2022}.}, days) to long-term (months and years; e.g. EX Lupi--type objects, or EXors, and FU Orionis--type objects, or FUors, which undergo eruptive accretion bursts; see, e.g., Fig.~3 by \citealp{Fischer2022}), and each of them produces distinctive photometric signatures across the wavelength domain. 
{Rubin LSST main survey (the Wide-Fast-Deep, WFD) will consist of a footprint of at least 18,000~deg$^2$. It must be uniformly covered to a median of 825 nominal 30~s visits per 9.6~deg$^2$ field, summed over all six filters (see \citealp{Bianco2022}).
This} baseline observing strategy is perfectly suited for the identification of EXors, whose long-term variability is characterized by brightness changes on the order of a few magnitudes over timescales of several months \citep[e.g.,][]{giannini2022}. However, a complementary, week-long period of dense (hourly) monitoring, fine-tuned to encompass all timescales of interest within and up to the typical stellar rotation rates of these young stars \citep[e.g.,][]{venuti2017}, is essential to extend the sensitivity of the main survey towards short-lived events. By combining these two components of the survey, we will be able to generate a properly populated light curve (LC) that samples all relevant time intervals to trace different star--disk phenomena. With this dual approach, we can exploit the full capability of Rubin LSST to discover and classify new YSO variable populations, particularly those undergoing disk accretion, and to characterize known objects (see also \citealp{Hambleton2022}). 

Up to now, systematic high-cadence observations of entire star-forming regions (SFRs), aimed at investigating short-term YSO variability, have been performed mostly from space, notably with the \emph{CoRoT} \citep{auvergne2009} and \emph{Kepler} \citep{borucki2010} spacecrafts, using a single broadband filter. Such surveys, targeting a handful of regions (e.g., \citealp{Alencar2010,Cody2014,Cody2018,Venuti2021,Cody2022}), have provided exquisite depictions of YSO variability, which allowed the implementation of statistical metrics to discriminate between distinct hour-to-month LC morphology classes for disk-bearing young stars, and automatically classify their respective behaviors across the time domain. However, these studies have shown that adding color information is essential for a coherent physical interpretation of the observed variability patterns. Moreover, the use of space-based facilities has so far greatly limited the list of suitable SFRs that could be targeted with those telescopes along their orbits. As both \textit{CoRoT} and \textit{Kepler} have been decommissioned, no opportunity for such high-cadence explorations of YSO variability is expected in the foreseeable future, and the currently operational TESS satellite \citep{ricker2015} is less suited for studies of SFRs due to the large pixel scale (21''/pixel) that hampers a proper identification of individual sources in potentially crowded environments.

Observations in the bluer regions of the visible spectrum are key to investigating YSO variability, as the blue bands are probes for the magnetospheric accretion process.
In classical T Tauri stars (CTTSs; low-mass YSOs undergoing mass accretion from the disk), blue band fluxes rise more during accretion events. The contemporaneous availability of red band fluxes allows us to disentangle accretion signatures from magnetic activity or extinction-induced variability on the basis of the shape of spectral energy distribution from the emitting region of the YSO \citep[e.g.,][]{vrba1993}. Furthermore, red-band (e.g., $r$) and blue-band (e.g.,~$u$) magnitudes can be combined in color-magnitude diagrams to identify the photometric cluster sequence traced by weak-line T Tauri stars (WTTSs; weakly or non-accreting YSOs; \citealp[e.g.,][]{richert2018}), while CTTS members can be discriminated upon their distinctive blueward shift in color with respect to the WTTS locus (see, e.g., Fig.~6 of \citealp{Venuti2014} for the NGC~2264 cluster).

In addition to characterizing high-amplitude variability from accreting CTTSs, LSST will also detect variability from non-accreting WTTSs due to their high levels of magnetic activity. Many WTTSs have large cool starspots that are rotationally modulated, producing periodic, $\sim$0.1 magnitude photometric variations on timescales of $\sim$1--20 days \citep[e.g.,][]{bouvier1993,herbst1994,venuti2015}. Brief ``white-light flares'' may also be seen \citep{stassun2006}, particularly among the fainter lower mass WTTSs. \looseness=-1

Our strategy for the investigation of YSOs and their variability allows us to fully exploit the capabilities offered by Rubin LSST during the whole duration of the survey operations. Indeed, both short-term variability, requiring the dense temporal coverage described here, and the long-term variability, mostly driven by EXor-like eruptive bursts, can be explored with this approach.
The unique capabilities of Rubin LSST will be complemented by archival data from high-precision and/or all-sky multi-band photometric surveys, like KELT \citep{pepper2007}, ASAS-SN \citep[e.g.,][]{jayasinghe2020}, Gaia \citep{gaia2022}, and ZTF \citep{bellm2019}. Such auxiliary datasets can be used for ``precovery" \citep{Yao2019}, that is, to check for historical variability to put the variations observed by Rubin Observatory into a larger context (Ustamujic et al. 2023a, in preparation). Furthermore, the aggregate temporal baseline covered by these surveys may enable the discovery of rarer, FUor--like outbursting events, characterized by longer (decades) timescales and more intense ($\gtrsim$5~magnitudes) variations \citep[e.g.,][]{kospal2021}.

In the following, we discuss our proposed short-term YSO variability monitoring and its impact on the wider Rubin LSST survey in terms of time investment and scientific gains by taking as reference the Carina Nebula. 
This region was identified as a starting target for our monitoring campaign because it guarantees a large number of sources (see \citealp{Townsley2011}, where 11,000 members are identified and a total population of up to 50,000--100,000 is estimated), and it is ideally placed for observations from Chile with Rubin LSST. Many members of the Carina Nebula complex, such as the Trumpler~15 cluster and unclustered dispersed WTTSs, are several million years old where the O stars have exploded as supernovae \citep{wang2011,feigelson2011,townsley2011b}. The richness and complexity of the stellar nurseries encompassed by the nebula has just been revealed in unprecedented detail by the very first images acquired with the NIRCam and MIRI cameras onboard the JWST \citep{greenhouse2019},
making it a perfect target to study the environmental feedback triggered by the radiation field from young, massive stars. 

Our paper is organized as follows: in Sect.~\ref{Metrics}, we present our framework for identification of short-term YSO variability behaviors with Rubin LSST; in Sect.~\ref{Results}, we compare our predictions with the output of different simulated runs for Rubin LSST and evaluate their performance with respect to our metrics; in Sect.~\ref{Discussion}, we discuss our proposed, comprehensive survey of YSO variability in representative SFRs, to be conducted along the entire Rubin LSST duration, to achieve a first systematic exploration of the early stellar evolution dynamics across a range of stellar masses, ages, and environments.

\section{Metrics description}
\label{Metrics}

To trace short-term variability phenomena in YSO populations belonging to distinct SFRs with Rubin LSST, we aim to draw on the space-borne experience and implement a cadence similar to the \emph{Kepler/K2} $\sim$30-minute ``long cadence''. Namely, we will sample the LCs of our targets with one point every 30 minutes over a 10 hour/night observing window for 7 consecutive nights in each of the following filters: $g$, $r$, $i$, and potentially\footnote{A lower number of visits are planned in the $u$-band with respect to the other Rubin LSST filters (see Table 2 by \citealp{Bianco2022}; see also Table 1 and Fig. 4 by \citealp{Ivezic2019}). Furthermore, the depth that will be reached in the $u$-band will be lower with respect to the other filters, therefore $u$-band data will not be available for faint sources.} $u$. The length of the observing window was selected so as to encompass the typical rotation rates measured for these young stars, which are believed to match the inner disk dynamical timescales as a result of the disk-locking mechanism enforced by the stellar magnetosphere \citep[e.g.,][]{rebull2022}. The total number of visits would then correspond to gathering 140 datapoints in one week across all filters. As mentioned in Sect.~\ref{sec:intro}, accretion events are most efficiently traced in the bluest filters available in Rubin LSST, such as $g$ or $u$\footnote{To be used under dark sky conditions.}, especially sensitive to the energetic emission from the accretion shock that is formed where magnetically-channeled, free-falling material from the disk impacts the star \citep[e.g.,][]{Gullbring1998}. The contemporaneous availability of redder optical filters, particularly $r$ and $i$, is pivotal to enable an estimation of stellar parameters such as spectral type and extinction, and to define the reference photospheric/chromospheric emission level above which accretion-related effects can be measured (\citealp[e.g.,][]{Venuti2014}; \citealp{Venuti2021}).

To simulate the implementation of our short-term YSO variability monitoring project with Rubin LSST, and therefore assess its feasibility and overall impact on the survey, we have worked in close collaboration with the Metrics Analysis Framework (MAF\footnote{See \citet{Bianco2022} for details.}) team, who developed a run of the Rubin LSST Operations Simulator (OpSim; \citealp{Naghib2019}) fine-tuned on our specific requirement for a dense temporal coverage of SFRs, starting with Carina Nebula as a testbed (\texttt{carina} OpSim)\footnote{See \url{http://astro-lsst-01.astro.washington.edu:8080/} for all available OpSim runs, including \texttt{carina}.}.
As noted in Sect.~\ref{sec:intro} and discussed in detail in Sect.~\ref{Discussion} (see also the White Paper \citealp{Bonito2018} and the Cadence Note by Bonito and Venuti et al. 2021\footnote{All Cadence Notes related to Rubin LSST survey strategy stored at \url{https://www.lsst.org/content/survey-cadence-notes-2021}.}), this observing run is conceived as the first step of a microsurvey\footnote{As explained in \citet{Bianco2022}, a Rubin microsurvey is defined as a specific observing campaign, distinct from the main LSST survey, that will require $<$3\% of the total survey time.} aimed at exploring the dynamics and evolution of young star-disk interaction as a function of intrinsic and external parameters across the early pre-main sequence. To achieve this goal, we plan to apply the same observing strategy on different SFRs one target each year during the implementation of the main Rubin LSST WFD survey.

In order to define the cadence and sampling criteria for identification of YSO short-term variability behaviors, we started from the light curves of young stars with variability patterns dominated by intense and unstable accretion activity \citep{Kulkarni2008}, manifested in short-lived luminosity bursts that arise repeatedly (recurrence timescales of days to weeks) and decay on timescales as short as several hours. Observational descriptions of such variability behaviors have been achieved over the last decade thanks to dedicated space-based monitoring campaigns of young stellar populations (\citealp{Stauffer2014}; \citealp{Cody2017}).
We focused in particular on the LCs of YSOs identified as busters in the open clusters NGC~2264 (monitored with \textit{CoRoT}; \citealp{Stauffer2014}, \citealp{Cody2014}) and NGC~6530 (with \textit{Kepler/K2}; \citealp{Venuti2021}). 

For each original LC (spanning a duration $\Delta t$ of $\sim$38 days in the case of \textit{CoRoT} time series, and $\sim$72 days in the case of \textit{K2} time series), we extracted a set of simulated LSST datasets by randomly generating an initial epoch $t_{init}$ comprised, within each LC, between observing times $t_1=0$ and $t_2=\Delta t - 7$ days. We then extracted a 7-day segment from the satellite time series starting at $t_{init}$, and retained 10 hours of data every 24 hours within that segment to reproduce the night/day alternation that will be present in Rubin LSST ground-based observations. We then considered alternative samplings in our simulated datasets: one point every 30 minutes (the same cadence as \textit{K2}), and lower cadences of one point every 45, 60, 90, 120, and 180 minutes. For each simulated sampling, we only selected those datapoints in the retained LC segments that match the required time spacing. At each iteration, we overlapped the original time series and the resampled segments to assess whether a burst detection would occur in the simulated LSST dataset. We defined a positive burst detection occurrence when the selected points in the light curve cover at least the full excursion of one original bursting event (from bottom to peak), as well as the out-of-burst luminosity level for comparison. We further defined a non-detection as an instance where the selected points only span the out-of-burst variability amplitudes in flux, and a potential burst detection as an instance where the selected points cover both the typical stellar luminosity level and a phase of anomalous brightening, but the recorded brightening does not extend beyond $\sim$1.5\,$\sigma$ above the typical luminosity level.

\begin{figure}
\centering
\includegraphics[width=0.47\textwidth]{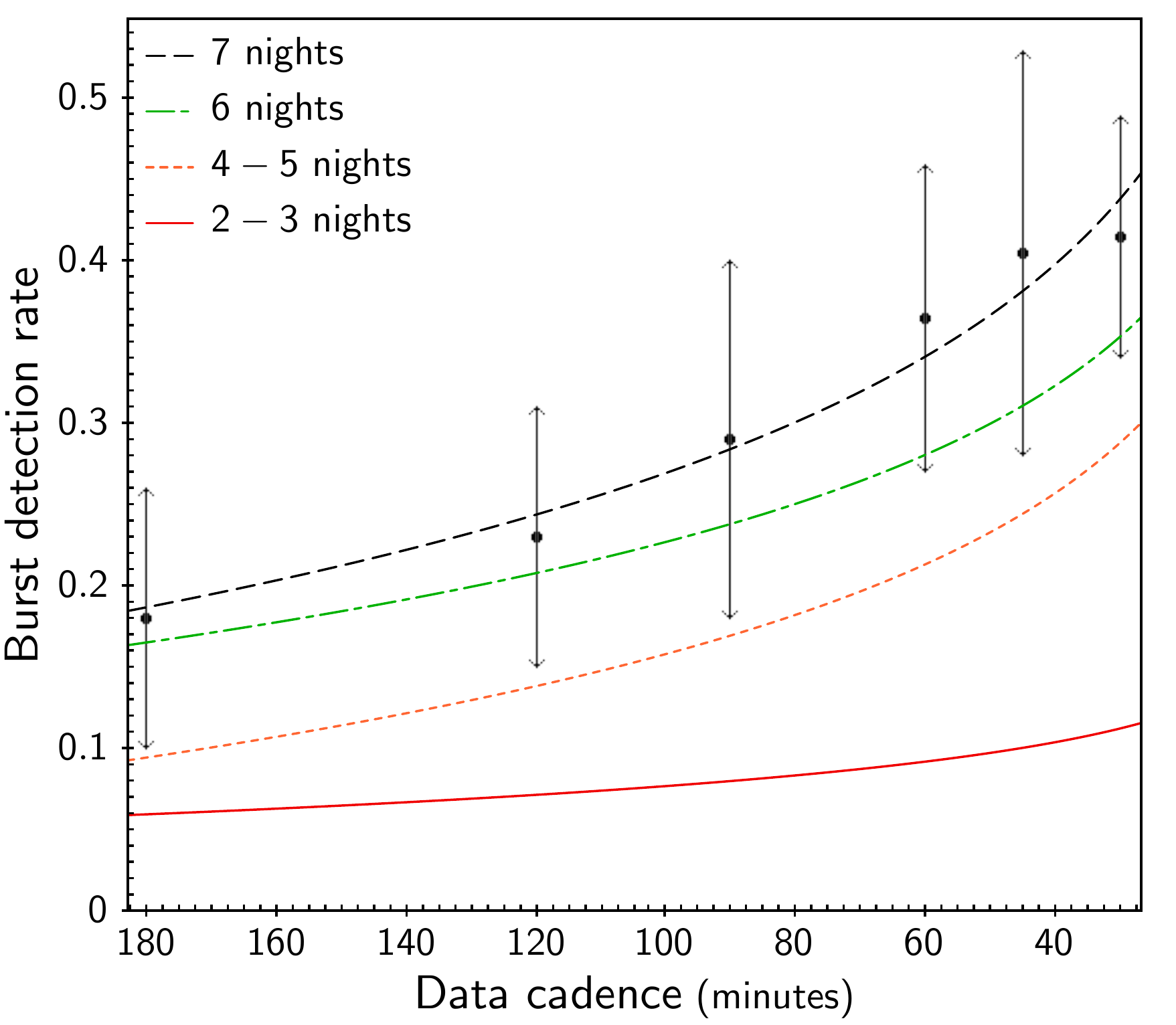}
\caption{Rate of accretion burst detection simulated for a high-cadence Rubin monitoring program of a given SFR, extending over one week. For each simulation, datapoints are assumed to be distributed uniformly and spaced in time by the amount on the $x$-axis. The black dots correspond to the average detection rates calculated from the simulation results assuming 7 consecutive observing days, with 10 hours of observations per night. The double arrows mark the statistical range of burst detection rates extracted from the simulated datasets around each average estimate. The dashed black line traces a logarithmic fit to the calculated detection rates to interpolate for intermediate cadences. The green (dash-dotted), orange (dotted), and red (solid) lines, labeled in the legend, illustrate how the trend of detection rate vs. observing cadence would change if one or more nights of observations during the week were lost to bad weather or other technical or scheduling issues. Based on \textit{CoRoT}/\textit{Kepler} observations of
YSOs in NGC 2264 and NGC 6530, respectively.} 
\label{fig:burst-detection}
\end{figure}

We collected the results of over 1,600 simulated sets comprising the six different cadences listed above, applied to a sample of 27 original bursting LCs. Gaps in the original time series were used to simulate the potential impact of missing nights (due to, e.g., time lost to bad weather during the program implementation at the Rubin Observatory) on our ability to infer a correct classification of YSO variability. Figure~\ref{fig:burst-detection} illustrates the projected detection rate of bursting events simulated for the Rubin LSST survey as a function of data cadence (which, in case of uniformly distributed observations, coincides with the duration of the longest segment of night not sampled by data relevant to YSO variability) and number of effective observing nights within one week. The statistical inferences from this analysis are discussed in Sect. \ref{Results}.

We then used the predictions illustrated in Fig.~\ref{fig:burst-detection} as a reference to evaluate the performance of different OpSim runs, and of the underlying assumptions in observing strategy, with respect to our science case of identifying short-lived variability phenomena in YSOs.
To conduct this comparison, we employed two different tools, which sift through the visits simulated in the OpSim for a given observing field throughout the duration of Rubin LSST and retain those that match specific positional and/or temporal criteria. More specifically, we adopted the metrics called \texttt{TransientAsciiMetric}\footnote{Original Python notebook available at \url{https://github.com/LSST-nonproject/sims_maf_contrib/blob/master/science/Transients/TransientAsciiMetric.ipynb}.} to scan the generic main survey OpSims and extract the number of observations expected for a given field in a given filter over seven consecutive days. We then adopted an analogous filtering routine, dubbed \texttt{Carina}\footnote{Original Python notebook available at \url{https://github.com/yoachim/21_Scratch/blob/main/carina_check/carina2.ipynb}.}, which conducts the same search as the \texttt{TransientAsciiMetric} on the \texttt{carina} OpSim developed by the MAF team to meet our scientific requirements. Results are presented in Sect. \ref{Results}.

\section{Results}
\label{Results}

As shown in Fig.~\ref{fig:burst-detection}, by using the \textit{K2} data of NGC~6530 and the \textit{CoRoT} data of NGC~2264, we estimated that the probability of detecting bursting events on a given object with erratic, accretion-driven variability would reach up to $45-50\%$ over a one-week long monitoring campaign using a 30- to 60-minute cadence. This plateauing value of the detection rate vs. data cadence trend is likely a reflection of the intrinsic recurrence timescales of short--term bursting behaviors in YSOs \citep{Cody2017} with respect to the prospected duration of the Rubin LSST high-cadence YSO monitoring survey: indeed, more than 40\% of the burster--like YSO variables identified with \textit{K2} exhibit burst repeat timescales of one week or longer. On the other hand, lower data cadences do affect the simulated burst detection rates beyond the impact of the monitoring duration. Indeed, the projected achievable burst detection rate would drop to less than $30\%$ with a 120-minute cadence ($30-40\%$ science loss), and to less than $25\%$ with a 180-minute cadence ($\sim 40-50\%$ science loss). 

Figure~\ref{fig:burst-detection} also illustrates the potential impact of having our monitoring series interrupted for one night or more, due to, for instance, bad weather or other scheduling conflicts. As shown on the diagram, across all explored cadences the estimated burst detection rate is not impacted significantly by a potential gap of a single night along the one-week monitoring window: the discrepancy between the average trends reconstructed from our simulations when assuming seven consecutive nights or six nights with a one-night gap in-between is smaller than (or consistent with) the computed statistical range of burst detection rates. On the other hand, more frequent gaps in the time series would have a substantial impact on our ability to identify burster stars. Limiting the observations to 4--5 out of the 7 nights  would cause a science loss (in terms of detected bursts) of 30--50\% across the observing cadence range in Fig.~\ref{fig:burst-detection}. Most importantly, were data to be taken only on 2--3 nights within the week (which still corresponds to a higher frequency of weekly visits than planned for any given field, and in single filters, as part of the WFD survey), the probability of detecting bursting events driven by the inner disk dynamics would become negligible (corresponding to a  science loss of 65--75\% with respect to the yield from the 7-day observing strategy at the matching cadence). 

{While we may miss the intermediate timescales, which can be explored from longer-term Rubin LSST observations (as investigated in Ustamujic et al., in preparation, using available datasets from large-scale surveys), existing studies of rotation rate distributions in diverse young stellar populations (see, e.g., \citealp{venuti2017}, for NGC 2264; \citealp{Roquette2017}, for Cygnus OB2; \citealp{Rebull2018}, for Upper Scorpius and rho Ophiuchus; \citealp{Rebull2020}, for Taurus) suggest that a consecutive 7 days-long window is statistically sufficient to explore the full rotational variability expected for at least 70\% of young stars. Furthermore, the selected range (7-days) here discussed would allow us to cover at least the half-period variability for the vast majority (95\%-98\%) of YSOs.}

To assess which observing strategies, among those implemented for different families of OpSims, would allow us to accomplish the discussed science goals, we compared our predictions shown in Fig.~\ref{fig:burst-detection} with the frequency of visits that would be obtained on a hypothetical SFR field, within one week and for a given filter, according to each of the OpSims below for the main survey:\looseness=-1
\begin{itemize}
    \item \texttt{baseline\_v2.1\_10yrs.db}, corresponding to the standard WFD strategy implemented throughout the entire Rubin LSST, with the exception of five select areas, called Deep Drilling Fields (DDFs), for which a denser observing cadence is adopted (totaling about 5\% of survey observations)\footnote{More details on the different observing strategies explored regarding the fraction of time to be spent on DDF coverage can be found at \url{https://pstn-051.lsst.io/PSTN-051.pdf}.};
    \item \texttt{ddf\_heavy\_nexp2\_v1.6\_10yrs.db}, an iteration of the WFD baseline survey with intensified coverage of the DDF regions, reaching over 13\% of the total survey observations; 
    \item \texttt{carina\_v2.2\_10yrs.db}, created ad hoc by the MAF team to include one week of intense SFR monitoring, as here described, during each year of the main survey. 
\end{itemize}

\begin{figure}
\includegraphics[width=0.47\textwidth]{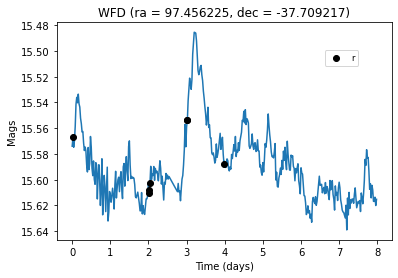}
\includegraphics[width=0.47\textwidth]{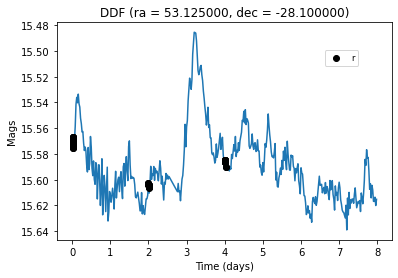}
\caption{Light curve of a YSO exhibiting a bursting behavior (blue curve; data from \textit{CoRoT}) with black points indicating the visits in the $r$-band that would be obtained during the most favorable week in the 10 years of survey, as selected with the \texttt{TransientAsciiMetric} applied to the OpSim \texttt{baseline\_v2.1\_10yrs}, if the object were alternatively located in a generic field to be surveyed with the WFD cadence (upper panel; ra = 97.456225, dec = -37.709217), or in a DDF region (lower panel; ra = 53.125, dec = -28.100).
\label{fig:baseline}}
\end{figure}

\begin{figure}
\includegraphics[width=0.47\textwidth]{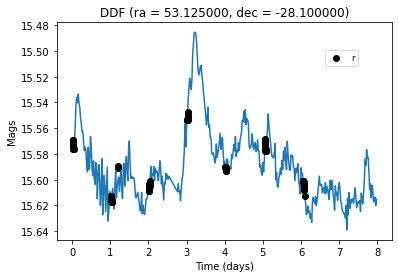}
\caption{Same as Fig. \ref{fig:baseline} (lower panel), reflecting the denser DDF observing cadence but implemented for the OpSim database \texttt{ddf\_heavy\_nexp2\_v1.6\_10yrs}. 
\label{fig:ddf-heavy}}
\end{figure}

\begin{figure}
\includegraphics[width=0.47\textwidth]{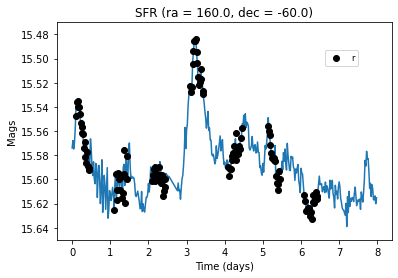}
\caption{Same as Fig.~\ref{fig:baseline}, but using the OpSim database \texttt{carina\_v2.2\_10yrs} and the \texttt{Carina} routine, for which a specific SFR (Carina Nebula; ra = 160.0, dec = -60.0) observing cadence was implemented following the sampling requirements discussed here.
\label{fig:carina-LC}}
\end{figure}

With the \texttt{TransientAsciiMetric}, we scanned, week by week, the entire database of simulated 10-year LSST observations corresponding to each of the listed OpSims, and extracted the typical number of expected visits on representative fields during a full week, along with the number of visits corresponding to the densest (most favorable) coverage for that field in a 7-day window along the simulated dataset. This search revealed that:
\begin{enumerate}[a)]
\item on a generic WFD field within the OpSim \texttt{baseline\_v2.1\_10yrs.db}, we found a maximum of $7$ points\footnote{It is worth noting that 7 points/week is a very unusual case for WFD single-filter coverage, as the mean value considering all the weeks in 10 years is close to 0 (i.e., less than one visit per week).} in $\sim$7 consecutive days over the 10 years of survey to populate the LC of a hypothetical YSO at that location, exemplified in Fig. \ref{fig:baseline} (upper panel) by using as input the LC of a burster in NGC~2264 that we used in our simulations;
\item if we adopt the sequence of observations obtained for the location of one of the DDFs, and assume that a similar sequence was obtained for the Carina region, then we would collect at most 62 datapoints (and typically $\sim$8 datapoints/week) with the OpSim \texttt{baseline\_v2.1\_10yrs.db} (Fig.~\ref{fig:baseline}, lower panel), clustered around very short time intervals and split in three blocks a few days apart;
\item for the same DDF, monitored as simulated in the OpSim \texttt{ddf\_heavy\_nexp2\_v1.6\_10yrs.db}, we retrieved a maximum of 143 points (and typically 24 points) over a span of 7 days in the $r$-band (see Fig. \ref{fig:ddf-heavy})\footnote{As not all DDF simulations have been repeated (MAF team, priv. comm.), we discuss here the results obtained for version 1.6.}, however the points would not provide a uniform sampling of each monitored night, but rather cover a very short time interval during each of the nights for which data would be available;
\item using the OpSim database \texttt{carina\_v2.2\_10yrs.db} and the \texttt{Carina} routine, we retrieve a frequency and cadence of observations that match the requirements to reconstruct short-term variability in YSOs (see also \citealp{Bonito2018} and the Cadence Note by Bonito and Venuti et al. 2021), as illustrated in Fig. \ref{fig:carina-LC} where the morphology of the bursting LC is adequately sampled with the expected visits.
\end{enumerate}

The case discussed in a) and shown in Fig.~\ref{fig:baseline} (upper panel), with typically zero and at most seven datapoints over four nights in one week, would fall outside (to the left) of the nightly data cadence range that is shown on the $x$-axis in Fig.~\ref{fig:burst-detection}, corresponding to a long effective cadence dictated by the actual fraction of night not covered by data. Along the $y$-axis of Fig.~\ref{fig:burst-detection}, possible observing runs under this OpSim would span the entire area delimited on top by the orange (dotted) line (datapoints taken on 4--5 nights within one week) and extending well below the red (solid) curve (which illustrates the case when datapoints are taken on 2--3 nights within one week). This would translate to an expected burst detection rate $\lesssim$~2\%, effectively preventing any robust identification of short-lived variability behaviors in a statistical sample of young stars. 

The case discussed in b) and illustrated in the lower panel of Fig.~\ref{fig:baseline}, in spite of the higher number of datapoints, would still fall below the long-cadence end of the orange (dotted) curve in Fig.~\ref{fig:burst-detection} (burst detection rate $\sim$2--3\%), because the sequence of expected visits for a given date only span a fraction ($\simeq$2\%) of the observing night, and data are gathered only on typically three nights of a consecutive week, in the best-case scenario. 

The case discussed in c) and shown in Fig.~\ref{fig:ddf-heavy} is even more emblematic. Despite attaining, in the most favorable simulated week, a number of visits that matches our ideal expectation for a 7-day run with a 30-minute cadence, this scenario would again correspond to a burst detection rate of only $\sim$2--10\% according to Fig.~\ref{fig:burst-detection} (extrapolation of the green dash-dotted line to intra-night data gaps equal to $\gtrsim$55--95\% of the night itself). Indeed, although repeated visits in this simulation are conducted daily, the very limited fraction of night that is actually sampled severely hampers our capability of identifying a bursting event over its full evolution. Only when both a dense, week-long coverage and a uniform sampling of the intra-night YSO variability are achieved (as discussed in d) and shown in Fig.~\ref{fig:carina-LC}), can the ability to recognize and categorize short-term phenomena be maximized\footnote{Additional plots can be found at \url{https://github.com/sbonito/Cadence-Note-YSO}.}. 

\section{Discussion and conclusions}
\label{Discussion}

We have here illustrated how it is possible to exploit the full capability of Rubin Observatory and explore the diverse physical processes at work in YSOs over a wide range of timescales. In particular, we have shown how a short ($\sim$week) period of dense observing cadence in multiple filters is critical to accurately reconstruct the inner disk dynamics around young stars and its short-lived manifestations, like luminosity bursts triggered by intense and discrete accretion events. When combined with the long--term, lower-cadence monitoring from the main LSST survey (which will naturally capture the larger-scale variations characteristic of EXors or FUors), this campaign will deliver unique science by mapping the intermediate stages between short-term and long-term variability dynamics in YSOs, thereby revealing any causal link between small-scale disk processes and large-scale eruptive behaviors. We have further shown that none of the standard WFD observing strategies under consideration for the Rubin LSST baseline and DDF surveys are able to provide the combination of dense cadence and homogeneous sampling over the inner disk timescales that are required to achieve our science goals. 

In collaboration with the MAF team, we have estimated that the 1-week period of high-cadence monitoring discussed here would require only about $\approx 0.02 \%$ of the total survey time for a single SFR, such as the Carina Nebula. This would amount to a total of $\approx 0.2 \%$ of Rubin observations if these focused campaigns were repeated once per year over the entire survey duration, each time targeting a different SFR (in order to achieve a census of YSO properties at different evolutionary stages and in different environments) or returning to the same SFR at a distance of years (to assess structural changes in the inner disk, which have been documented to occur on timescales of three years or less, with switches from well-organized to chaotic dust distributions; \citealp{sousa2016}). Therefore, by investing a small fraction of Rubin survey time, we can address key science questions for protoplanetary disk and early stellar evolution (regarding, e.g., the regulation of mass and angular momentum transfer, the mechanisms triggering outbursts in YSOs, and the interplay between accretion and magnetic/coronal activity) that would otherwise be precluded. Moreover, the impact of a well-sampled dataset as proposed here would extend well beyond the topics discussed so far, to encompass, for instance, the early angular momentum evolution of young stars (by enabling homogeneous measurements of their rotation rates).

As noted in Sect.~\ref{sec:intro}, we plan to begin our microsurvey with a Carina Nebula pilot project. By extending the campaign to additional SFRs, and by combining these short periods of intense monitoring with WFD observations gathered as part of the Galactic Plane minisurvey (Cadence Note by Street et al. 2021; Prisinzano et al. 2022, submitted) across the entire Rubin LSST duration, we will be able to build a comprehensive, self-consistent picture of variability in YSOs. We stress that employing Rubin Observatory for both the short-term and the long-term components of these studies provides a unique advantage over using other facilities: only Rubin can guarantee the spatial resolution, consistent filter prescription, and photometric stability required to conduct a coherent comparative analysis of flux variations recorded for the same sources on timescales of hours through years, with a precision better than $\sim 0.01$~mag (i.e., on the order of the smallest variability amplitudes of interest; \citealp{costigan2014,Venuti2014}).

The synoptic view of accretion and variability properties that Rubin LSST will provide for hundreds of young stars in a single snapshot will be pivotal to guide complementary ground-based campaigns, aimed at gathering supporting data, like spectroscopy, for accurate determination of the stellar parameters.
Different instruments for spectroscopic investigation, including currently available and future facilities like FLAMES \citep{pasquini2000}, X-shooter \citep{vernet2011}, SoXS \citep{schipani2018}, CRIRES+ \citep{dorn2014}, 4MOST \citep{dejong2022}, and WEAVE \citep{dalton2020}, can be used in concert with Rubin to build a comprehensive photometric and spectroscopic variability atlas of young star-disk systems. 
Indeed, an intensive campaign focused on short-term variability as described here, extending over just one week at a time, will allow us to envision a coordinated spectroscopic program for the entire duration of the high-cadence monitoring series -- important to spectroscopically confirm accretion processes (e.g., from measuring broad emission lines produced by the accelerated gas in the accretion columns; e.g., \citealp{bonito2013, bonito2020}), and to constrain theoretical models.   

\begin{deluxetable}{c c c c c}
\centerwidetable
\caption{List of star-forming regions to be targeted during the dedicated Rubin Observatory microsurvey, and their respective ages, distances, typical amounts of interstellar extinction, and number of currently known cluster members. \label{tab:regions}}
\startdata
\tablehead{\colhead{Region} & \colhead{Age} & \colhead{Distance} & \colhead{$A_V$} & \colhead{$N_\mathrm{stars}$}\\
 & [Myr] & [kpc] & [mag] & }
Carina Nebula & 1--6 & 2.3 & 2.25 & 7300+\\
NGC~6530 (Lagoon) & 1--2 & 1.3 & 1.65 & 2000+\\
Orion Nebula Cluster & 1--3 & 0.4 & 1.5 & 1500+\\
NGC~6611 (Eagle) & 1--3 & 1.8 & 2.8 & 2500+\\
NGC~2264 & 3--5 & 0.8 & 0.4 & 1100+
\enddata
\tablecomments{The listed $N_\mathrm{stars}$ for all regions are extracted from \citet{feigelson2013}. Additional references include: \citet{smith2008} for the Carina Nebula; \citet{tothill2008} and \citet{prisinzano2019} for NGC~6530; \citet{muench2008}, \citet{dario2010} for the Orion Nebula Cluster; \citet{oliveira2008} for NGC~6611; \citet{dahm2008} for NGC~2264.}
\end{deluxetable}

The different SFRs\footnote{More maps stored at \url{http://astro-lsst-01.astro.washington.edu:8080/} and \url{https://github.com/LSST-TVSSC/software_tools/blob/main/footprint_maps/bonito_sfr_map_g.png}.} that we plan to target as part of our microsurvey are listed in Table~\ref{tab:regions}. These targets will enable detailed investigations of how the processes responsible for the observed YSO variability evolve over the protoplanetary disk lifetimes, and of how their dynamics is impacted by different ambient conditions. 
An early-time start of this microsurvey within the LSST schedule, beginning with observations of the Carina Nebula, will be crucial to set the stage for exploration of all SFRs. Indeed, while for a few of the listed regions (e.g., NGC~2264, \citealp{Venuti2014}; NGC~6530, \citealp{Venuti2021}) some short-term monitoring data in similar filters already exist and a single Rubin visit could be sufficient to extend the baseline for inner disk stability studies, for the other targets multiple Rubin visits, separated by a few years, will be required. Following our approach of observing only one cluster per year, this objective cannot be achieved for all of our regions during the prospected LSST duration unless the microsurvey starts promptly upon beginning of Rubin science operations. Moreover, an early scientific analysis of the Carina Nebula data, with its rich trove of members, will allow us to develop reliable photometric tracers of young variable populations in a combination of Rubin LSST filters, which can then be used to conduct a blind search of the entire Rubin LSST field to identify new pre-main sequence populations. This step will be critical to realize the full discovery potential of the LSST survey beyond the $\sim500$~pc radius within which most star formation studies have traditionally been confined.

Results from this project will, on one hand, provide improved physical insight to constrain existing and future magnetohydrodynamic models of young star--disk systems \citep[e.g.,][]{romanova2004,romanova2012,zanni2009,orlando2010,zanni2013,Bonito2014}, and on the other hand, also inform the development of realistic 3D sketch models that describe the complex structure of these objects (Ustamujic et al. 2023b, in preparation). Printed versions of these 3D renderings, together with their visualization in Virtual Reality (VR) experiences connected to the physical processes investigated (see \citealp{orlando2019}), will allow us to pursue a more inclusive scientific environment by increasing the accessibility of Rubin LSST’s results for Blind and Visually Impaired (BVI) researchers. Moreover, thanks to their interactive and immersive nature \citep[e.g.,][]{jarrett2021}, these tools will also prove invaluable in assisting the broader community of astronomers in their exploration and  research of the complex astrophysical systems that can be investigated with Vera C. Rubin Observatory LSST. 

\begin{acknowledgements}

This paper was created in the nursery of the Rubin LSST  Transient and Variable Stars (TVS) Science Collaboration (SC)\footnote{\url{https://lsst-tvssc.github.io/}}.
The authors acknowledge the support of the Vera C. Rubin LSST TVS SC that provided opportunities for collaboration and exchange of ideas and knowledge.
We would like to acknowledge the support that John Stauffer (1952-2021) gave to this project since the preparation of the White Paper \cite{Bonito2018}.
The authors are thankful for the support provided by the Vera C. Rubin Observatory MAF team in the creation and implementation of MAFs.
The authors acknowledge the support of the LSST Corporation that enabled the organization of many workshops and hackathons throughout the cadence optimization process.
This work was supported by the Preparing for Astrophysics with LSST Program, funded by the Heising Simons Foundation through grant 2021-2975, and administered by Las Cumbres Observatory (``Young stellar objects and their variability with Rubin LSST: combining observations and 3D models for a more inclusive Science" -- In memory of Fabio Bonito, PI: R. Bonito and L. Venuti).
R.B. and A.C.G. acknowledge financial support from the project PRIN-INAF 2019 ``Spectroscopically Tracing the Disk Dispersal Evolution". 
A.C.G. was also supported by PRIN-INAF MAIN-STREAM 2017 ``Protoplanetary disks seen through the eyes of new-generation instruments''.
L.V. is supported by the National Aeronautics and Space Administration (NASA) under grant No. 80NSSC21K0633 issued through the NNH20ZDA001N Astrophysics Data Analysis Program (ADAP). 
L.V. is grateful to INAF--Osservatorio Astronomico di Palermo for hosting her during her visit in June 2022 for Rubin LSST research. L.V. also warmly acknowledges the Centre for the Subatomic Structure of Matter for their hospitality during her stays at the University of Adelaide as a Visitor in 2022, when parts of this manuscript were written. 

\end{acknowledgements}

\bibliography{bonitovenuti}{}

\begin{thebibliography}{}
\expandafter\ifx\csname natexlab\endcsname\relax\def\natexlab#1{#1}\fi
\providecommand{\url}[1]{\href{#1}{#1}}
\providecommand{\dodoi}[1]{doi:~\href{http://doi.org/#1}{\nolinkurl{#1}}}
\providecommand{\doeprint}[1]{\href{http://ascl.net/#1}{\nolinkurl{http://ascl.net/#1}}}
\providecommand{\doarXiv}[1]{\href{https://arxiv.org/abs/#1}{\nolinkurl{https://arxiv.org/abs/#1}}}

\bibitem[{{Alencar} {et~al.}(2010){Alencar}, {Teixeira}, {Guimar{\~a}es},
  {McGinnis}, {Gameiro}, {Bouvier}, {Aigrain}, {Flaccomio}, \&
  {Favata}}]{Alencar2010}
{Alencar}, S.~H.~P., {Teixeira}, P.~S., {Guimar{\~a}es}, M.~M., {et~al.} 2010,
  \aap, 519, A88, \dodoi{10.1051/0004-6361/201014184}

\bibitem[{{Auvergne} {et~al.}(2009){Auvergne}, {Bodin}, {Boisnard}, {Buey},
  {Chaintreuil}, {Epstein}, {Jouret}, {Lam-Trong}, {Levacher}, {Magnan},
  {Perez}, {Plasson}, {Plesseria}, {Peter}, {Steller}, {Tiph{\`e}ne}, {Baglin},
  {Agogu{\'e}}, {Appourchaux}, {Barbet}, {Beaufort}, {Bellenger}, {Berlin},
  {Bernardi}, {Blouin}, {Boumier}, {Bonneau}, {Briet}, {Butler}, {Cautain},
  {Chiavassa}, {Costes}, {Cuvilho}, {Cunha-Parro}, {de Oliveira Fialho},
  {Decaudin}, {Defise}, {Djalal}, {Docclo}, {Drummond}, {Dupuis}, {Exil},
  {Faur{\'e}}, {Gaboriaud}, {Gamet}, {Gavalda}, {Grolleau}, {Gueguen},
  {Guivarc'h}, {Guterman}, {Hasiba}, {Huntzinger}, {Hustaix}, {Imbert},
  {Jeanville}, {Johlander}, {Jorda}, {Journoud}, {Karioty}, {Kerjean},
  {Lafond}, {Lapeyrere}, {Landiech}, {Larqu{\'e}}, {Laudet}, {Le Merrer},
  {Leporati}, {Leruyet}, {Levieuge}, {Llebaria}, {Martin}, {Mazy}, {Mesnager},
  {Michel}, {Moalic}, {Monjoin}, {Naudet}, {Neukirchner}, {Nguyen-Kim},
  {Ollivier}, {Orcesi}, {Ottacher}, {Oulali}, {Parisot}, {Perruchot},
  {Piacentino}, {Pinheiro da Silva}, {Platzer}, {Pontet}, {Pradines},
  {Quentin}, {Rohbeck}, {Rolland}, {Rollenhagen}, {Romagnan}, {Russ}, {Samadi},
  {Schmidt}, {Schwartz}, {Sebbag}, {Smit}, {Sunter}, {Tello}, {Toulouse},
  {Ulmer}, {Vandermarcq}, {Vergnault}, {Wallner}, {Waultier}, \&
  {Zanatta}}]{auvergne2009}
{Auvergne}, M., {Bodin}, P., {Boisnard}, L., {et~al.} 2009, \aap, 506, 411,
  \dodoi{10.1051/0004-6361/200810860}

\bibitem[{{Bellm} {et~al.}(2022){Bellm}, {Burke}, {Coughlin}, {Andreoni},
  {Raiteri}, \& {Bonito}}]{Bellm2022}
{Bellm}, E.~C., {Burke}, C.~J., {Coughlin}, M.~W., {et~al.} 2022, \apjs, 258,
  13, \dodoi{10.3847/1538-4365/ac4602}

\bibitem[{{Bellm} {et~al.}(2019){Bellm}, {Kulkarni}, {Graham}, {Dekany},
  {Smith}, {Riddle}, {Masci}, {Helou}, {Prince}, {Adams}, {Barbarino},
  {Barlow}, {Bauer}, {Beck}, {Belicki}, {Biswas}, {Blagorodnova}, {Bodewits},
  {Bolin}, {Brinnel}, {Brooke}, {Bue}, {Bulla}, {Burruss}, {Cenko}, {Chang},
  {Connolly}, {Coughlin}, {Cromer}, {Cunningham}, {De}, {Delacroix}, {Desai},
  {Duev}, {Eadie}, {Farnham}, {Feeney}, {Feindt}, {Flynn}, {Franckowiak},
  {Frederick}, {Fremling}, {Gal-Yam}, {Gezari}, {Giomi}, {Goldstein},
  {Golkhou}, {Goobar}, {Groom}, {Hacopians}, {Hale}, {Henning}, {Ho}, {Hover},
  {Howell}, {Hung}, {Huppenkothen}, {Imel}, {Ip}, {Ivezi{\'c}}, {Jackson},
  {Jones}, {Juric}, {Kasliwal}, {Kaspi}, {Kaye}, {Kelley}, {Kowalski},
  {Kramer}, {Kupfer}, {Landry}, {Laher}, {Lee}, {Lin}, {Lin}, {Lunnan},
  {Giomi}, {Mahabal}, {Mao}, {Miller}, {Monkewitz}, {Murphy}, {Ngeow},
  {Nordin}, {Nugent}, {Ofek}, {Patterson}, {Penprase}, {Porter}, {Rauch},
  {Rebbapragada}, {Reiley}, {Rigault}, {Rodriguez}, {van Roestel}, {Rusholme},
  {van Santen}, {Schulze}, {Shupe}, {Singer}, {Soumagnac}, {Stein}, {Surace},
  {Sollerman}, {Szkody}, {Taddia}, {Terek}, {Van Sistine}, {van Velzen},
  {Vestrand}, {Walters}, {Ward}, {Ye}, {Yu}, {Yan}, \& {Zolkower}}]{bellm2019}
{Bellm}, E.~C., {Kulkarni}, S.~R., {Graham}, M.~J., {et~al.} 2019, \pasp, 131,
  018002, \dodoi{10.1088/1538-3873/aaecbe}

\bibitem[{{Bianco} {et~al.}(2022){Bianco}, {Ivezi{\'c}}, {Jones}, {Graham},
  {Marshall}, {Saha}, {Strauss}, {Yoachim}, {Ribeiro}, {Anguita}, {Bauer},
  {Bauer}, {Bellm}, {Blum}, {Brandt}, {Brough}, {Catelan}, {Clarkson},
  {Connolly}, {Gawiser}, {Gizis}, {Hlo{\v{z}}ek}, {Kaviraj}, {Liu}, {Lochner},
  {Mahabal}, {Mandelbaum}, {McGehee}, {Neilsen}, {Olsen}, {Peiris}, {Rhodes},
  {Richards}, {Ridgway}, {Schwamb}, {Scolnic}, {Shemmer}, {Slater}, {Slosar},
  {Smartt}, {Strader}, {Street}, {Trilling}, {Verma}, {Vivas}, {Wechsler}, \&
  {Willman}}]{Bianco2022}
{Bianco}, F.~B., {Ivezi{\'c}}, {\v{Z}}., {Jones}, R.~L., {et~al.} 2022, \apjs,
  258, 1, \dodoi{10.3847/1538-4365/ac3e72}

\bibitem[{{Bonito} {et~al.}(2014){Bonito}, {Orlando}, {Argiroffi}, {Miceli},
  {Peres}, {Matsakos}, {Stehle}, \& {Ibgui}}]{Bonito2014}
{Bonito}, R., {Orlando}, S., {Argiroffi}, C., {et~al.} 2014, \apjl, 795, L34,
  \dodoi{10.1088/2041-8205/795/2/L34}

\bibitem[{{Bonito} {et~al.}(2013){Bonito}, {Prisinzano}, {Guarcello}, \&
  {Micela}}]{bonito2013}
{Bonito}, R., {Prisinzano}, L., {Guarcello}, M.~G., \& {Micela}, G. 2013, \aap,
  556, A108, \dodoi{10.1051/0004-6361/201321405}

\bibitem[{{Bonito} {et~al.}(2018){Bonito}, {Hartigan}, {Venuti}, {Guarcello},
  {Prisinzano}, {Argiroffi}, {Messina}, {Johns-Krull}, {Feigelson}, {Stauffer},
  {Giannini}, {Antoniucci}, {Sciortino}, {Micela}, {Pillitteri}, {Fedele},
  {Podio}, {Damiani}, {McGehee}, {Street}, {Gizis}, {Sacco}, {Magrini},
  {Flaccomio}, {Orlando}, {Miceli}, {Stelzer}, {Fuchs}, {Chen}, {Pikuz},
  {Frasca}, {Biazzo}, {Codella}, {Pastorello}, {Alcala'}, {Covino}, {Bianchi},
  \& {Nisini}}]{Bonito2018}
{Bonito}, R., {Hartigan}, P., {Venuti}, L., {et~al.} 2018, arXiv e-prints,
  arXiv:1812.03135.
\newblock \doarXiv{1812.03135}

\bibitem[{{Bonito} {et~al.}(2020){Bonito}, {Prisinzano}, {Venuti}, {Damiani},
  {Micela}, {Sacco}, {Traven}, {Biazzo}, {Sbordone}, {Masseron}, {Zwitter},
  {Gonneau}, {Bayo}, {Roccatagliata}, {Randich}, {Vink}, {Jofre}, {Flaccomio},
  {Magrini}, {Carraro}, {Morbidelli}, {Frasca}, {Monaco}, {Rigliaco}, {Worley},
  {Hourihane}, {Gilmore}, {Franciosini}, {Lewis}, \& {Koposov}}]{bonito2020}
{Bonito}, R., {Prisinzano}, L., {Venuti}, L., {et~al.} 2020, \aap, 642, A56,
  \dodoi{10.1051/0004-6361/202037942}

\bibitem[{{Borucki} {et~al.}(2010){Borucki}, {Koch}, {Basri}, {Batalha},
  {Brown}, {Caldwell}, {Caldwell}, {Christensen-Dalsgaard}, {Cochran},
  {DeVore}, {Dunham}, {Dupree}, {Gautier}, {Geary}, {Gilliland}, {Gould},
  {Howell}, {Jenkins}, {Kondo}, {Latham}, {Marcy}, {Meibom}, {Kjeldsen},
  {Lissauer}, {Monet}, {Morrison}, {Sasselov}, {Tarter}, {Boss}, {Brownlee},
  {Owen}, {Buzasi}, {Charbonneau}, {Doyle}, {Fortney}, {Ford}, {Holman},
  {Seager}, {Steffen}, {Welsh}, {Rowe}, {Anderson}, {Buchhave}, {Ciardi},
  {Walkowicz}, {Sherry}, {Horch}, {Isaacson}, {Everett}, {Fischer}, {Torres},
  {Johnson}, {Endl}, {MacQueen}, {Bryson}, {Dotson}, {Haas}, {Kolodziejczak},
  {Van Cleve}, {Chandrasekaran}, {Twicken}, {Quintana}, {Clarke}, {Allen},
  {Li}, {Wu}, {Tenenbaum}, {Verner}, {Bruhweiler}, {Barnes}, \&
  {Prsa}}]{borucki2010}
{Borucki}, W.~J., {Koch}, D., {Basri}, G., {et~al.} 2010, Science, 327, 977,
  \dodoi{10.1126/science.1185402}

\bibitem[{{Bouvier} {et~al.}(1993){Bouvier}, {Cabrit}, {Fernandez}, {Martin},
  \& {Matthews}}]{bouvier1993}
{Bouvier}, J., {Cabrit}, S., {Fernandez}, M., {Martin}, E.~L., \& {Matthews},
  J.~M. 1993, \aap, 272, 176

\bibitem[{{Cody} \& {Hillenbrand}(2018)}]{Cody2018}
{Cody}, A.~M., \& {Hillenbrand}, L.~A. 2018, \aj, 156, 71,
  \dodoi{10.3847/1538-3881/aacead}

\bibitem[{{Cody} {et~al.}(2017){Cody}, {Hillenbrand}, {David}, {Carpenter},
  {Everett}, \& {Howell}}]{Cody2017}
{Cody}, A.~M., {Hillenbrand}, L.~A., {David}, T.~J., {et~al.} 2017, \apj, 836,
  41, \dodoi{10.3847/1538-4357/836/1/41}

\bibitem[{{Cody} {et~al.}(2022){Cody}, {Hillenbrand}, \& {Rebull}}]{Cody2022}
{Cody}, A.~M., {Hillenbrand}, L.~A., \& {Rebull}, L.~M. 2022, \aj, 163, 212,
  \dodoi{10.3847/1538-3881/ac5b73}

\bibitem[{{Cody} {et~al.}(2014){Cody}, {Stauffer}, {Baglin}, {Micela},
  {Rebull}, {Flaccomio}, {Morales-Calder{\'o}n}, {Aigrain}, {Bouvier},
  {Hillenbrand}, {Gutermuth}, {Song}, {Turner}, {Alencar}, {Zwintz},
  {Plavchan}, {Carpenter}, {Findeisen}, {Carey}, {Terebey}, {Hartmann},
  {Calvet}, {Teixeira}, {Vrba}, {Wolk}, {Covey}, {Poppenhaeger}, {G{\"u}nther},
  {Forbrich}, {Whitney}, {Affer}, {Herbst}, {Hora}, {Barrado}, {Holtzman},
  {Marchis}, {Wood}, {Medeiros Guimar{\~a}es}, {Lillo Box}, {Gillen},
  {McQuillan}, {Espaillat}, {Allen}, {D'Alessio}, \& {Favata}}]{Cody2014}
{Cody}, A.~M., {Stauffer}, J., {Baglin}, A., {et~al.} 2014, \aj, 147, 82,
  \dodoi{10.1088/0004-6256/147/4/82}

\bibitem[{{Costigan} {et~al.}(2014){Costigan}, {Vink}, {Scholz}, {Ray}, \&
  {Testi}}]{costigan2014}
{Costigan}, G., {Vink}, J.~S., {Scholz}, A., {Ray}, T., \& {Testi}, L. 2014,
  \mnras, 440, 3444, \dodoi{10.1093/mnras/stu529}

\bibitem[{{Da Rio} {et~al.}(2010){Da Rio}, {Robberto}, {Soderblom}, {Panagia},
  {Hillenbrand}, {Palla}, \& {Stassun}}]{dario2010}
{Da Rio}, N., {Robberto}, M., {Soderblom}, D.~R., {et~al.} 2010, \apj, 722,
  1092, \dodoi{10.1088/0004-637X/722/2/1092}

\bibitem[{{Dahm}(2008)}]{dahm2008}
{Dahm}, S.~E. 2008, in Handbook of Star Forming Regions, Volume I, ed.
  B.~{Reipurth}, Vol.~4, 966

\bibitem[{{Dalton} {et~al.}(2020){Dalton}, {Trager}, {Abrams}, {Bonifacio},
  {Aguerri}, {Vallenari}, {Bishop}, {Middleton}, {Benn}, {Dee}, {Mignot},
  {Lewis}, {Pragt}, {Pico}, {Walton}, {Rey}, {Allende Prieto}, {Lhom{\'e}},
  {Balcells}, {Terrett}, {Brock}, {Ridings}, {Skvar{\v{c}}}, {Verheijen},
  {Steele}, {Stuik}, {Kroes}, {Tromp}, {Kragt}, {Lesman}, {Mottram}, {Bates},
  {Gribbin}, {Burgal}, {Herreros}, {Delgado}, {Martin}, {Cano}, {Navarro},
  {Irwin}, {Peralta de Arriba}, {O'Mahoney}, {Bianco}, {Moleinezhad}, {ter
  Horst}, {Molinari}, {Lodi}, {Guerra}, {Baruffalo}, {Carrasco}, {Farcas},
  {Schallig}, {Hughes}, {Hill}, {Smith}, {Drew}, {Poggianti}, {Iovino},
  {Pieri}, {Jin}, {Dominguez Palmero}, {Fari{\~n}a}, {Mart{\'\i}n}, {Worley},
  {Murphy}, {Guest}, {Morris}, {Elswijk}, {de Haan}, {Hanenburg}, {Salasnich},
  {Mayya}, {Izazaga-P{\'e}rez}, {Gafton}, {Caffau}, {Horville}, {Paz
  Chinch{\'o}n}, {Falcon-Barosso}, {G{\"a}nsicke}, {San Juan}, \&
  {Hernandez}}]{dalton2020}
{Dalton}, G., {Trager}, S., {Abrams}, D.~C., {et~al.} 2020, in Society of
  Photo-Optical Instrumentation Engineers (SPIE) Conference Series, Vol. 11447,
  Society of Photo-Optical Instrumentation Engineers (SPIE) Conference Series,
  1144714, \dodoi{10.1117/12.2561067}

\bibitem[{{de Jong} {et~al.}(2022){de Jong}, {Bellido-Tirado}, {Brynnel},
  {Ezzati Amini}, {Frey}, {F{\"u}{\ss}lein}, {G{\"a}bler}, {Giannone}, {Johl},
  {Kuba}, {Lemke}, {Micheva}, {Saviauk}, {Steinmetz}, {Walcher}, {Winkler},
  {Lind}, {Loveday}, {Feltzing}, {McMahon}, {Mainieri}, {Pirard}, {Bensby},
  {Bergemann}, {Chiappini}, {Christlieb}, {Cioni}, {Comparat}, {Driver},
  {Hook}, {Irwin}, {Kneib}, {Liske}, {Merloni}, {Minchev}, {Richard},
  {Starkenburg}, {Sullivan}, {Worley}, {Gaessler}, {Laurent}, {Pragt},
  {Remillieux}, {Rothmaier}, {Smedley}, {Stilz}, {Walton}, {Alexander},
  {Church}, {Croom}, {Davies}, {Heneka}, {Kacharov}, {Knoche}, {Kordopatis},
  {Krumpe}, {Martell}, {Norberg}, {Pelisoli}, {Sharma}, {Storm}, \&
  {Tempel}}]{dejong2022}
{de Jong}, R.~S., {Bellido-Tirado}, O., {Brynnel}, J.~G., {et~al.} 2022, in
  Society of Photo-Optical Instrumentation Engineers (SPIE) Conference Series,
  Vol. 12184, Ground-based and Airborne Instrumentation for Astronomy IX, ed.
  C.~J. {Evans}, J.~J. {Bryant}, \& K.~{Motohara}, 1218414,
  \dodoi{10.1117/12.2628965}

\bibitem[{{Dorn} {et~al.}(2014){Dorn}, {Anglada-Escude}, {Baade}, {Bristow},
  {Follert}, {Gojak}, {Grunhut}, {Hatzes}, {Heiter}, {Hilker}, {Ives}, {Jung},
  {K{\"a}ufl}, {Kerber}, {Klein}, {Lizon}, {Lockhart}, {L{\"o}winger},
  {Marquart}, {Oliva}, {Origlia}, {Pasquini}, {Paufique}, {Piskunov}, {Pozna},
  {Reiners}, {Smette}, {Smoker}, {Seemann}, {Stempels}, \&
  {Valenti}}]{dorn2014}
{Dorn}, R.~J., {Anglada-Escude}, G., {Baade}, D., {et~al.} 2014, The Messenger,
  156, 7

\bibitem[{{Feigelson} {et~al.}(2011){Feigelson}, {Getman}, {Townsley}, {Broos},
  {Povich}, {Garmire}, {King}, {Montmerle}, {Preibisch}, {Smith}, {Stassun},
  {Wang}, {Wolk}, \& {Zinnecker}}]{feigelson2011}
{Feigelson}, E.~D., {Getman}, K.~V., {Townsley}, L.~K., {et~al.} 2011, \apjs,
  194, 9, \dodoi{10.1088/0067-0049/194/1/9}

\bibitem[{{Feigelson} {et~al.}(2013){Feigelson}, {Townsley}, {Broos}, {Busk},
  {Getman}, {King}, {Kuhn}, {Naylor}, {Povich}, {Baddeley}, {Bate},
  {Indebetouw}, {Luhman}, {McCaughrean}, {Pittard}, {Pudritz}, {Sills}, {Song},
  \& {Wadsley}}]{feigelson2013}
{Feigelson}, E.~D., {Townsley}, L.~K., {Broos}, P.~S., {et~al.} 2013, \apjs,
  209, 26, \dodoi{10.1088/0067-0049/209/2/26}

\bibitem[{{Fischer} {et~al.}(2022){Fischer}, {Hillenbrand}, {Herczeg},
  {Johnstone}, {K{\'o}sp{\'a}l}, \& {Dunham}}]{Fischer2022}
{Fischer}, W.~J., {Hillenbrand}, L.~A., {Herczeg}, G.~J., {et~al.} 2022, arXiv
  e-prints, arXiv:2203.11257.
\newblock \doarXiv{2203.11257}

\bibitem[{{Gaia Collaboration} {et~al.}(2022){Gaia Collaboration}, {Vallenari},
  {Brown}, {Prusti}, {de Bruijne}, {Arenou}, {Babusiaux}, {Biermann},
  {Creevey}, {Ducourant}, {Evans}, {Eyer}, {Guerra}, {Hutton}, {Jordi},
  {Klioner}, {Lammers}, {Lindegren}, {Luri}, {Mignard}, {Panem}, {Pourbaix},
  {Randich}, {Sartoretti}, {Soubiran}, {Tanga}, {Walton}, {Bailer-Jones},
  {Bastian}, {Drimmel}, {Jansen}, {Katz}, {Lattanzi}, {van Leeuwen}, {Bakker},
  {Cacciari}, {Casta{\~n}eda}, {De Angeli}, {Fabricius}, {Fouesneau},
  {Fr{\'e}mat}, {Galluccio}, {Guerrier}, {Heiter}, {Masana}, {Messineo},
  {Mowlavi}, {Nicolas}, {Nienartowicz}, {Pailler}, {Panuzzo}, {Riclet}, {Roux},
  {Seabroke}, {Sordo{\o}rcit}, {Th{\'e}venin}, {Gracia-Abril}, {Portell},
  {Teyssier}, {Altmann}, {Andrae}, {Audard}, {Bellas-Velidis}, {Benson},
  {Berthier}, {Blomme}, {Burgess}, {Busonero}, {Busso}, {C{\'a}novas}, {Carry},
  {Cellino}, {Cheek}, {Clementini}, {Damerdji}, {Davidson}, {de Teodoro},
  {Nu{\~n}ez Campos}, {Delchambre}, {Dell'Oro}, {Esquej},
  {Fern{\'a}ndez-Hern{\'a}ndez}, {Fraile}, {Garabato}, {Garc{\'\i}a-Lario},
  {Gosset}, {Haigron}, {Halbwachs}, {Hambly}, {Harrison}, {Hern{\'a}ndez},
  {Hestroffer}, {Hodgkin}, {Holl}, {Jan{\ss}en}, {Jevardat de Fombelle},
  {Jordan}, {Krone-Martins}, {Lanzafame}, {L{\"o}ffler}, {Marchal}, {Marrese},
  {Moitinho}, {Muinonen}, {Osborne}, {Pancino}, {Pauwels}, {Recio-Blanco},
  {Reyl{\'e}}, {Riello}, {Rimoldini}, {Roegiers}, {Rybizki}, {Sarro}, {Siopis},
  {Smith}, {Sozzetti}, {Utrilla}, {van Leeuwen}, {Abbas}, {{\'A}brah{\'a}m},
  {Abreu Aramburu}, {Aerts}, {Aguado}, {Ajaj}, {Aldea-Montero}, {Altavilla},
  {{\'A}lvarez}, {Alves}, {Anders}, {Anderson}, {Anglada Varela}, {Antoja},
  {Baines}, {Baker}, {Balaguer-N{\'u}{\~n}ez}, {Balbinot}, {Balog}, {Barache},
  {Barbato}, {Barros}, {Barstow}, {Bartolom{\'e}}, {Bassilana}, {Bauchet},
  {Becciani}, {Bellazzini}, {Berihuete}, {Bernet}, {Bertone}, {Bianchi},
  {Binnenfeld}, {Blanco-Cuaresma}, {Blazere}, {Boch}, {Bombrun}, {Bossini},
  {Bouquillon}, {Bragaglia}, {Bramante}, {Breedt}, {Bressan}, {Brouillet},
  {Brugaletta}, {Bucciarelli}, {Burlacu}, {Butkevich}, {Buzzi}, {Caffau},
  {Cancelliere}, {Cantat-Gaudin}, {Carballo}, {Carlucci}, {Carnerero},
  {Carrasco}, {Casamiquela}, {Castellani}, {Castro-Ginard}, {Chaoul},
  {Charlot}, {Chemin}, {Chiaramida}, {Chiavassa}, {Chornay}, {Comoretto},
  {Contursi}, {Cooper}, {Cornez}, {Cowell}, {Crifo}, {Cropper}, {Crosta},
  {Crowley}, {Dafonte}, {Dapergolas}, {David}, {David}, {de Laverny}, {De
  Luise}, {De March}, {De Ridder}, {de Souza}, {de Torres}, {del Peloso}, {del
  Pozo}, {Delbo}, {Delgado}, {Delisle}, {Demouchy}, {Dharmawardena}, {Di
  Matteo}, {Diakite}, {Diener}, {Distefano}, {Dolding}, {Edvardsson}, {Enke},
  {Fabre}, {Fabrizio}, {Faigler}, {Fedorets}, {Fernique}, {Fienga}, {Figueras},
  {Fournier}, {Fouron}, {Fragkoudi}, {Gai}, {Garcia-Gutierrez},
  {Garcia-Reinaldos}, {Garc{\'\i}a-Torres}, {Garofalo}, {Gavel}, {Gavras},
  {Gerlach}, {Geyer}, {Giacobbe}, {Gilmore}, {Girona}, {Giuffrida}, {Gomel},
  {Gomez}, {Gonz{\'a}lez-N{\'u}{\~n}ez}, {Gonz{\'a}lez-Santamar{\'\i}a},
  {Gonz{\'a}lez-Vidal}, {Granvik}, {Guillout}, {Guiraud},
  {Guti{\'e}rrez-S{\'a}nchez}, {Guy}, {Hatzidimitriou}, {Hauser}, {Haywood},
  {Helmer}, {Helmi}, {Sarmiento}, {Hidalgo}, {Hilger}, {H{\l}adczuk}, {Hobbs},
  {Holland}, {Huckle}, {Jardine}, {Jasniewicz}, {Jean-Antoine Piccolo},
  {Jim{\'e}nez-Arranz}, {Jorissen}, {Juaristi Campillo}, {Julbe}, {Karbevska},
  {Kervella}, {Khanna}, {Kontizas}, {Kordopatis}, {Korn}, {K{\'o}sp{\'a}l},
  {Kostrzewa-Rutkowska}, {Kruszy{\'n}ska}, {Kun}, {Laizeau}, {Lambert},
  {Lanza}, {Lasne}, {Le Campion}, {Lebreton}, {Lebzelter}, {Leccia}, {Leclerc},
  {Lecoeur-Taibi}, {Liao}, {Licata}, {Lindstr{\o}m}, {Lister}, {Livanou},
  {Lobel}, {Lorca}, {Loup}, {Madrero Pardo}, {Magdaleno Romeo}, {Managau},
  {Mann}, {Manteiga}, {Marchant}, {Marconi}, {Marcos}, {Marcos Santos},
  {Mar{\'\i}n Pina}, {Marinoni}, {Marocco}, {Marshall}, {Polo},
  {Mart{\'\i}n-Fleitas}, {Marton}, {Mary}, {Masip}, {Massari},
  {Mastrobuono-Battisti}, {Mazeh}, {McMillan}, {Messina}, {Michalik}, {Millar},
  {Mints}, {Molina}, {Molinaro}, {Moln{\'a}r}, {Monari}, {Mongui{\'o}},
  {Montegriffo}, {Montero}, {Mor}, {Mora}, {Morbidelli}, {Morel}, {Morris},
  {Muraveva}, {Murphy}, {Musella}, {Nagy}, {Noval}, {Oca{\~n}a}, {Ogden},
  {Ordenovic}, {Osinde}, {Pagani}, {Pagano}, {Palaversa}, {Palicio},
  {Pallas-Quintela}, {Panahi}, {Payne-Wardenaar}, {Pe{\~n}alosa Esteller},
  {Penttil{\"a}}, {Pichon}, {Piersimoni}, {Pineau}, {Plachy}, {Plum}, {Poggio},
  {Pr{\v{s}}a}, {Pulone}, {Racero}, {Ragaini}, {Rainer}, {Raiteri}, {Rambaux},
  {Ramos}, {Ramos-Lerate}, {Re Fiorentin}, {Regibo}, {Richards}, {Rios Diaz},
  {Ripepi}, {Riva}, {Rix}, {Rixon}, {Robichon}, {Robin}, {Robin}, {Roelens},
  {Rogues}, {Rohrbasser}, {Romero-G{\'o}mez}, {Rowell}, {Royer}, {Ruz Mieres},
  {Rybicki}, {Sadowski}, {S{\'a}ez N{\'u}{\~n}ez}, {Sagrist{\`a} Sell{\'e}s},
  {Sahlmann}, {Salguero}, {Samaras}, {Sanchez Gimenez}, {Sanna},
  {Santove{\~n}a}, {Sarasso}, {Schultheis}, {Sciacca}, {Segol}, {Segovia},
  {S{\'e}gransan}, {Semeux}, {Shahaf}, {Siddiqui}, {Siebert}, {Siltala},
  {Silvelo}, {Slezak}, {Slezak}, {Smart}, {Snaith}, {Solano}, {Solitro},
  {Souami}, {Souchay}, {Spagna}, {Spina}, {Spoto}, {Steele},
  {Steidelm{\"u}ller}, {Stephenson}, {S{\"u}veges}, {Surdej}, {Szabados},
  {Szegedi-Elek}, {Taris}, {Taylo}, {Teixeira}, {Tolomei}, {Tonello}, {Torra},
  {Torra}, {Torralba Elipe}, {Trabucchi}, {Tsounis}, {Turon}, {Ulla}, {Unger},
  {Vaillant}, {van Dillen}, {van Reeven}, {Vanel}, {Vecchiato}, {Viala},
  {Vicente}, {Voutsinas}, {Weiler}, {Wevers}, {Wyrzykowski}, {Yoldas}, {Yvard},
  {Zhao}, {Zorec}, {Zucker}, \& {Zwitter}}]{gaia2022}
{Gaia Collaboration}, {Vallenari}, A., {Brown}, A.~G.~A., {et~al.} 2022, arXiv
  e-prints, arXiv:2208.00211.
\newblock \doarXiv{2208.00211}

\bibitem[{{Giannini} {et~al.}(2022){Giannini}, {Giunta}, {Gangi}, {Carini},
  {Lorenzetti}, {Antoniucci}, {Caratti o Garatti}, {Cassar{\'a}}, {Nisini},
  {Rossi}, {Testa}, \& {Vitali}}]{giannini2022}
{Giannini}, T., {Giunta}, A., {Gangi}, M., {et~al.} 2022, \apj, 929, 129,
  \dodoi{10.3847/1538-4357/ac5a49}

\bibitem[{Greenhouse(2019)}]{greenhouse2019}
Greenhouse, M. 2019, in 2019 IEEE Aerospace Conference, 1--13,
  \dodoi{10.1109/AERO.2019.8742209}

\bibitem[{{Gullbring} {et~al.}(1998){Gullbring}, {Hartmann}, {Brice{\~n}o}, \&
  {Calvet}}]{Gullbring1998}
{Gullbring}, E., {Hartmann}, L., {Brice{\~n}o}, C., \& {Calvet}, N. 1998, \apj,
  492, 323, \dodoi{10.1086/305032}

\bibitem[{{Hambleton} {et~al.}(2022){Hambleton}, {Bianco}, {Street}, {Bell},
  {Buckley}, {Graham}, {Hernitschek}, {Lund}, {Mason}, {Pepper}, {Prsa},
  {Rabus}, {Raiteri}, {Szabo}, {Szkody}, {Andreoni}, {Antoniucci},
  {Balmaverde}, {Bellm}, {Bonito}, {Bono}, {Botticella}, {Brocato}, {Bucar
  Bricman}, {Cappellaro}, {Carnerero}, {Chornock}, {Clarke}, {Cowperthwaite},
  {Cucchiara}, {D'Ammando}, {Dage}, {Dall'Ora}, {Davenport}, {de Martino}, {de
  Somma}, {Di Criscienzo}, {Di Stefano}, {Drout}, {Fabrizio}, {Fiorentino},
  {Gandhi}, {Garofalo}, {Giannini}, {Gomboc}, {Greggio}, {Hartigan},
  {Hundertmark}, {Johnson}, {Johnson}, {Jurkic}, {Khakpash}, {Leccia}, {Li},
  {Magurno}, {Malanchev}, {Marconi}, {Margutti}, {Marinoni}, {Mauron},
  {Molinaro}, {Moller}, {Moniez}, {Muraveva}, {Musella}, {Ngeow}, {Pastorello},
  {Petrecca}, {Piranomonte}, {Ragosta}, {Reguitti}, {Righi}, {Ripepi}, {Rivera
  Sandoval}, {Stassun}, {Stroh}, {Terreran}, {Trimble}, {Tsapras}, {van
  Velzen}, {Venuti}, \& {Vink}}]{Hambleton2022}
{Hambleton}, K.~M., {Bianco}, F.~B., {Street}, R., {et~al.} 2022, arXiv
  e-prints, arXiv:2208.04499.
\newblock \doarXiv{2208.04499}

\bibitem[{{Herbst} {et~al.}(1994){Herbst}, {Herbst}, {Grossman}, \&
  {Weinstein}}]{herbst1994}
{Herbst}, W., {Herbst}, D.~K., {Grossman}, E.~J., \& {Weinstein}, D. 1994, \aj,
  108, 1906, \dodoi{10.1086/117204}

\bibitem[{{Ivezi{\'c}} {et~al.}(2019){Ivezi{\'c}}, {Kahn}, {Tyson}, {Abel},
  {Acosta}, {Allsman}, {Alonso}, {AlSayyad}, {Anderson}, {Andrew}, {Angel},
  {Angeli}, {Ansari}, {Antilogus}, {Araujo}, {Armstrong}, {Arndt}, {Astier},
  {Aubourg}, {Auza}, {Axelrod}, {Bard}, {Barr}, {Barrau}, {Bartlett}, {Bauer},
  {Bauman}, {Baumont}, {Bechtol}, {Bechtol}, {Becker}, {Becla}, {Beldica},
  {Bellavia}, {Bianco}, {Biswas}, {Blanc}, {Blazek}, {Blandford}, {Bloom},
  {Bogart}, {Bond}, {Booth}, {Borgland}, {Borne}, {Bosch}, {Boutigny},
  {Brackett}, {Bradshaw}, {Brandt}, {Brown}, {Bullock}, {Burchat}, {Burke},
  {Cagnoli}, {Calabrese}, {Callahan}, {Callen}, {Carlin}, {Carlson},
  {Chandrasekharan}, {Charles-Emerson}, {Chesley}, {Cheu}, {Chiang}, {Chiang},
  {Chirino}, {Chow}, {Ciardi}, {Claver}, {Cohen-Tanugi}, {Cockrum}, {Coles},
  {Connolly}, {Cook}, {Cooray}, {Covey}, {Cribbs}, {Cui}, {Cutri}, {Daly},
  {Daniel}, {Daruich}, {Daubard}, {Daues}, {Dawson}, {Delgado}, {Dellapenna},
  {de Peyster}, {de Val-Borro}, {Digel}, {Doherty}, {Dubois},
  {Dubois-Felsmann}, {Durech}, {Economou}, {Eifler}, {Eracleous}, {Emmons},
  {Fausti Neto}, {Ferguson}, {Figueroa}, {Fisher-Levine}, {Focke}, {Foss},
  {Frank}, {Freemon}, {Gangler}, {Gawiser}, {Geary}, {Gee}, {Geha}, {Gessner},
  {Gibson}, {Gilmore}, {Glanzman}, {Glick}, {Goldina}, {Goldstein}, {Goodenow},
  {Graham}, {Gressler}, {Gris}, {Guy}, {Guyonnet}, {Haller}, {Harris},
  {Hascall}, {Haupt}, {Hernandez}, {Herrmann}, {Hileman}, {Hoblitt}, {Hodgson},
  {Hogan}, {Howard}, {Huang}, {Huffer}, {Ingraham}, {Innes}, {Jacoby}, {Jain},
  {Jammes}, {Jee}, {Jenness}, {Jernigan}, {Jevremovi{\'c}}, {Johns}, {Johnson},
  {Johnson}, {Jones}, {Juramy-Gilles}, {Juri{\'c}}, {Kalirai}, {Kallivayalil},
  {Kalmbach}, {Kantor}, {Karst}, {Kasliwal}, {Kelly}, {Kessler}, {Kinnison},
  {Kirkby}, {Knox}, {Kotov}, {Krabbendam}, {Krughoff}, {Kub{\'a}nek},
  {Kuczewski}, {Kulkarni}, {Ku}, {Kurita}, {Lage}, {Lambert}, {Lange},
  {Langton}, {Le Guillou}, {Levine}, {Liang}, {Lim}, {Lintott}, {Long},
  {Lopez}, {Lotz}, {Lupton}, {Lust}, {MacArthur}, {Mahabal}, {Mandelbaum},
  {Markiewicz}, {Marsh}, {Marshall}, {Marshall}, {May}, {McKercher}, {McQueen},
  {Meyers}, {Migliore}, {Miller}, {Mills}, {Miraval}, {Moeyens}, {Moolekamp},
  {Monet}, {Moniez}, {Monkewitz}, {Montgomery}, {Morrison}, {Mueller},
  {Muller}, {Mu{\~n}oz Arancibia}, {Neill}, {Newbry}, {Nief}, {Nomerotski},
  {Nordby}, {O'Connor}, {Oliver}, {Olivier}, {Olsen}, {O'Mullane}, {Ortiz},
  {Osier}, {Owen}, {Pain}, {Palecek}, {Parejko}, {Parsons}, {Pease},
  {Peterson}, {Peterson}, {Petravick}, {Libby Petrick}, {Petry},
  {Pierfederici}, {Pietrowicz}, {Pike}, {Pinto}, {Plante}, {Plate}, {Plutchak},
  {Price}, {Prouza}, {Radeka}, {Rajagopal}, {Rasmussen}, {Regnault}, {Reil},
  {Reiss}, {Reuter}, {Ridgway}, {Riot}, {Ritz}, {Robinson}, {Roby}, {Roodman},
  {Rosing}, {Roucelle}, {Rumore}, {Russo}, {Saha}, {Sassolas}, {Schalk},
  {Schellart}, {Schindler}, {Schmidt}, {Schneider}, {Schneider}, {Schoening},
  {Schumacher}, {Schwamb}, {Sebag}, {Selvy}, {Sembroski}, {Seppala}, {Serio},
  {Serrano}, {Shaw}, {Shipsey}, {Sick}, {Silvestri}, {Slater}, {Smith},
  {Smith}, {Sobhani}, {Soldahl}, {Storrie-Lombardi}, {Stover}, {Strauss},
  {Street}, {Stubbs}, {Sullivan}, {Sweeney}, {Swinbank}, {Szalay}, {Takacs},
  {Tether}, {Thaler}, {Thayer}, {Thomas}, {Thornton}, {Thukral}, {Tice},
  {Trilling}, {Turri}, {Van Berg}, {Vanden Berk}, {Vetter}, {Virieux},
  {Vucina}, {Wahl}, {Walkowicz}, {Walsh}, {Walter}, {Wang}, {Wang}, {Warner},
  {Wiecha}, {Willman}, {Winters}, {Wittman}, {Wolff}, {Wood-Vasey}, {Wu},
  {Xin}, {Yoachim}, \& {Zhan}}]{Ivezic2019}
{Ivezi{\'c}}, {\v{Z}}., {Kahn}, S.~M., {Tyson}, J.~A., {et~al.} 2019, \apj,
  873, 111, \dodoi{10.3847/1538-4357/ab042c}

\bibitem[{{Jarrett} {et~al.}(2021){Jarrett}, {Comrie}, {Marchetti},
  {Sivitilli}, {Macfarlane}, {Vitello}, {Becciani}, {Taylor}, {van der Hulst},
  {Serra}, {Katz}, \& {Cluver}}]{jarrett2021}
{Jarrett}, T.~H., {Comrie}, A., {Marchetti}, L., {et~al.} 2021, Astronomy and
  Computing, 37, 100502, \dodoi{10.1016/j.ascom.2021.100502}

\bibitem[{{Jayasinghe} {et~al.}(2020){Jayasinghe}, {Stanek}, {Kochanek},
  {Shappee}, {Holoien}, {Thompson}, {Prieto}, {Dong}, {Pawlak}, {Pejcha},
  {Shields}, {Pojmanski}, {Otero}, {Hurst}, {Britt}, \&
  {Will}}]{jayasinghe2020}
{Jayasinghe}, T., {Stanek}, K.~Z., {Kochanek}, C.~S., {et~al.} 2020, \mnras,
  491, 13, \dodoi{10.1093/mnras/stz2711}

\bibitem[{{K{\'o}sp{\'a}l} {et~al.}(2021){K{\'o}sp{\'a}l}, {Cruz-S{\'a}enz de
  Miera}, {White}, {{\'A}brah{\'a}m}, {Chen}, {Csengeri}, {Dong}, {Dunham},
  {Feh{\'e}r}, {Green}, {Hashimoto}, {Henning}, {Hogerheijde}, {Kudo}, {Liu},
  {Takami}, \& {Vorobyov}}]{kospal2021}
{K{\'o}sp{\'a}l}, {\'A}., {Cruz-S{\'a}enz de Miera}, F., {White}, J.~A.,
  {et~al.} 2021, \apjs, 256, 30, \dodoi{10.3847/1538-4365/ac0f09}

\bibitem[{{Kulkarni} \& {Romanova}(2008)}]{Kulkarni2008}
{Kulkarni}, A.~K., \& {Romanova}, M.~M. 2008, \mnras, 386, 673,
  \dodoi{10.1111/j.1365-2966.2008.13094.x}

\bibitem[{{Muench} {et~al.}(2008){Muench}, {Getman}, {Hillenbrand}, \&
  {Preibisch}}]{muench2008}
{Muench}, A., {Getman}, K., {Hillenbrand}, L., \& {Preibisch}, T. 2008, in
  Handbook of Star Forming Regions, Volume I, ed. B.~{Reipurth}, Vol.~4, 483

\bibitem[{{Naghib} {et~al.}(2019){Naghib}, {Yoachim}, {Vanderbei}, {Connolly},
  \& {Jones}}]{Naghib2019}
{Naghib}, E., {Yoachim}, P., {Vanderbei}, R.~J., {Connolly}, A.~J., \& {Jones},
  R.~L. 2019, \aj, 157, 151, \dodoi{10.3847/1538-3881/aafece}

\bibitem[{{Oliveira}(2008)}]{oliveira2008}
{Oliveira}, J.~M. 2008, in Handbook of Star Forming Regions, Volume II, ed.
  B.~{Reipurth}, Vol.~5, 599

\bibitem[{{Orlando} {et~al.}(2019){Orlando}, {Pillitteri}, {Bocchino},
  {Daricello}, \& {Leonardi}}]{orlando2019}
{Orlando}, S., {Pillitteri}, I., {Bocchino}, F., {Daricello}, L., \&
  {Leonardi}, L. 2019, Research Notes of the American Astronomical Society, 3,
  176, \dodoi{10.3847/2515-5172/ab5966}

\bibitem[{{Orlando} {et~al.}(2010){Orlando}, {Sacco}, {Argiroffi}, {Reale},
  {Peres}, \& {Maggio}}]{orlando2010}
{Orlando}, S., {Sacco}, G.~G., {Argiroffi}, C., {et~al.} 2010, \aap, 510, A71,
  \dodoi{10.1051/0004-6361/200913565}

\bibitem[{{Pasquini} {et~al.}(2000){Pasquini}, {Avila}, {Allaert}, {Ballester},
  {Biereichel}, {Buzzoni}, {Cavadore}, {Dekker}, {Delabre}, {Ferraro}, {Hill},
  {Kaufer}, {Kotzlowski}, {Lizon}, {Longinotti}, {Moureau}, {Palsa}, \&
  {Zaggia}}]{pasquini2000}
{Pasquini}, L., {Avila}, G., {Allaert}, E., {et~al.} 2000, in Society of
  Photo-Optical Instrumentation Engineers (SPIE) Conference Series, Vol. 4008,
  Optical and IR Telescope Instrumentation and Detectors, ed. M.~{Iye} \& A.~F.
  {Moorwood}, 129--140, \dodoi{10.1117/12.395491}

\bibitem[{{Pepper} {et~al.}(2007){Pepper}, {Pogge}, {DePoy}, {Marshall},
  {Stanek}, {Stutz}, {Poindexter}, {Siverd}, {O'Brien}, {Trueblood}, \&
  {Trueblood}}]{pepper2007}
{Pepper}, J., {Pogge}, R.~W., {DePoy}, D.~L., {et~al.} 2007, \pasp, 119, 923,
  \dodoi{10.1086/521836}

\bibitem[{{Prisinzano} {et~al.}(2019){Prisinzano}, {Damiani}, {Kalari},
  {Jeffries}, {Bonito}, {Micela}, {Wright}, {Jackson}, {Tognelli}, {Guarcello},
  {Vink}, {Klutsch}, {Jim{\'e}nez-Esteban}, {Roccatagliata},
  {Tautvai{\v{s}}ien{\.{e}}}, {Gilmore}, {Randich}, {Alfaro}, {Flaccomio},
  {Koposov}, {Lanzafame}, {Pancino}, {Bergemann}, {Carraro}, {Franciosini},
  {Frasca}, {Gonneau}, {Hourihane}, {Jofr{\'e}}, {Lewis}, {Magrini}, {Monaco},
  {Morbidelli}, {Sacco}, {Worley}, \& {Zaggia}}]{prisinzano2019}
{Prisinzano}, L., {Damiani}, F., {Kalari}, V., {et~al.} 2019, \aap, 623, A159,
  \dodoi{10.1051/0004-6361/201834870}

\bibitem[{{Rebull} {et~al.}(2020){Rebull}, {Stauffer}, {Cody}, {Hillenbrand},
  {Bouvier}, {Roggero}, \& {David}}]{Rebull2020}
{Rebull}, L.~M., {Stauffer}, J.~R., {Cody}, A.~M., {et~al.} 2020, \aj, 159,
  273, \dodoi{10.3847/1538-3881/ab893c}

\bibitem[{{Rebull} {et~al.}(2018){Rebull}, {Stauffer}, {Cody}, {Hillenbrand},
  {David}, \& {Pinsonneault}}]{Rebull2018}
---. 2018, \aj, 155, 196, \dodoi{10.3847/1538-3881/aab605}

\bibitem[{{Rebull} {et~al.}(2022){Rebull}, {Stauffer}, {Hillenbrand}, {Cody},
  {Kruse}, \& {Powell}}]{rebull2022}
{Rebull}, L.~M., {Stauffer}, J.~R., {Hillenbrand}, L.~A., {et~al.} 2022, \aj,
  164, 80, \dodoi{10.3847/1538-3881/ac75f1}

\bibitem[{{Richert} {et~al.}(2018){Richert}, {Getman}, {Feigelson}, {Kuhn},
  {Broos}, {Povich}, {Bate}, \& {Garmire}}]{richert2018}
{Richert}, A.~J.~W., {Getman}, K.~V., {Feigelson}, E.~D., {et~al.} 2018,
  \mnras, 477, 5191, \dodoi{10.1093/mnras/sty949}

\bibitem[{{Ricker} {et~al.}(2015){Ricker}, {Winn}, {Vanderspek}, {Latham},
  {Bakos}, {Bean}, {Berta-Thompson}, {Brown}, {Buchhave}, {Butler}, {Butler},
  {Chaplin}, {Charbonneau}, {Christensen-Dalsgaard}, {Clampin}, {Deming},
  {Doty}, {De Lee}, {Dressing}, {Dunham}, {Endl}, {Fressin}, {Ge}, {Henning},
  {Holman}, {Howard}, {Ida}, {Jenkins}, {Jernigan}, {Johnson}, {Kaltenegger},
  {Kawai}, {Kjeldsen}, {Laughlin}, {Levine}, {Lin}, {Lissauer}, {MacQueen},
  {Marcy}, {McCullough}, {Morton}, {Narita}, {Paegert}, {Palle}, {Pepe},
  {Pepper}, {Quirrenbach}, {Rinehart}, {Sasselov}, {Sato}, {Seager},
  {Sozzetti}, {Stassun}, {Sullivan}, {Szentgyorgyi}, {Torres}, {Udry}, \&
  {Villasenor}}]{ricker2015}
{Ricker}, G.~R., {Winn}, J.~N., {Vanderspek}, R., {et~al.} 2015, Journal of
  Astronomical Telescopes, Instruments, and Systems, 1, 014003,
  \dodoi{10.1117/1.JATIS.1.1.014003}

\bibitem[{{Romanova} {et~al.}(2004){Romanova}, {Ustyugova}, {Koldoba}, \&
  {Lovelace}}]{romanova2004}
{Romanova}, M.~M., {Ustyugova}, G.~V., {Koldoba}, A.~V., \& {Lovelace},
  R.~V.~E. 2004, \apj, 610, 920, \dodoi{10.1086/421867}

\bibitem[{{Romanova} {et~al.}(2012){Romanova}, {Ustyugova}, {Koldoba}, \&
  {Lovelace}}]{romanova2012}
---. 2012, \mnras, 421, 63, \dodoi{10.1111/j.1365-2966.2011.20055.x}

\bibitem[{{Roquette} {et~al.}(2017){Roquette}, {Bouvier}, {Alencar}, {Vaz}, \&
  {Guarcello}}]{Roquette2017}
{Roquette}, J., {Bouvier}, J., {Alencar}, S.~H.~P., {Vaz}, L.~P.~R., \&
  {Guarcello}, M.~G. 2017, \aap, 603, A106, \dodoi{10.1051/0004-6361/201630337}

\bibitem[{{Schipani} {et~al.}(2018){Schipani}, {Campana}, {Claudi},
  {K{\"a}ufl}, {Accardo}, {Aliverti}, {Baruffolo}, {Ben Ami}, {Biondi},
  {Brucalassi}, {Capasso}, {Cosentino}, {D'Alessio}, {D'Avanzo}, {Hershko},
  {Gardiol}, {Kuncarayacti}, {Munari}, {Rubin}, {Scuderi}, {Vitali},
  {Achr{\'e}n}, {Araiza-Duran}, {Arcavi}, {Bianco}, {Cappellaro}, {Colapietro},
  {Della Valle}, {Diner}, {D'Orsi}, {Fantinel}, {Fynbo}, {Gal-Yam}, {Genoni},
  {Hirvonen}, {Kotilainen}, {Kumar}, {Landoni}, {Lehti}, {Li Causi},
  {Loreggia}, {Marafatto}, {Mattila}, {Pariani}, {Pignata}, {Rappaport},
  {Ricci}, {Riva}, {Salasnich}, {Zanmar Sanchez}, {Smartt}, \&
  {Turatto}}]{schipani2018}
{Schipani}, P., {Campana}, S., {Claudi}, R., {et~al.} 2018, in Society of
  Photo-Optical Instrumentation Engineers (SPIE) Conference Series, Vol. 10702,
  Ground-based and Airborne Instrumentation for Astronomy VII, ed. C.~J.
  {Evans}, L.~{Simard}, \& H.~{Takami}, 107020F, \dodoi{10.1117/12.2307349}

\bibitem[{{Smith} \& {Brooks}(2008)}]{smith2008}
{Smith}, N., \& {Brooks}, K.~J. 2008, in Handbook of Star Forming Regions,
  Volume II, ed. B.~{Reipurth}, Vol.~5, 138

\bibitem[{{Sousa} {et~al.}(2016){Sousa}, {Alencar}, {Bouvier}, {Stauffer},
  {Venuti}, {Hillenbrand}, {Cody}, {Teixeira}, {Guimar{\~a}es}, {McGinnis},
  {Rebull}, {Flaccomio}, {F{\"u}r{\'e}sz}, {Micela}, \& {Gameiro}}]{sousa2016}
{Sousa}, A.~P., {Alencar}, S.~H.~P., {Bouvier}, J., {et~al.} 2016, \aap, 586,
  A47, \dodoi{10.1051/0004-6361/201526599}

\bibitem[{{Stassun} {et~al.}(2006){Stassun}, {van den Berg}, {Feigelson}, \&
  {Flaccomio}}]{stassun2006}
{Stassun}, K.~G., {van den Berg}, M., {Feigelson}, E., \& {Flaccomio}, E. 2006,
  \apj, 649, 914, \dodoi{10.1086/506422}

\bibitem[{{Stauffer} {et~al.}(2014){Stauffer}, {Cody}, {Baglin}, {Alencar},
  {Rebull}, {Hillenbrand}, {Venuti}, {Turner}, {Carpenter}, {Plavchan},
  {Findeisen}, {Carey}, {Terebey}, {Morales-Calder{\'o}n}, {Bouvier}, {Micela},
  {Flaccomio}, {Song}, {Gutermuth}, {Hartmann}, {Calvet}, {Whitney}, {Barrado},
  {Vrba}, {Covey}, {Herbst}, {Furesz}, {Aigrain}, \& {Favata}}]{Stauffer2014}
{Stauffer}, J., {Cody}, A.~M., {Baglin}, A., {et~al.} 2014, \aj, 147, 83,
  \dodoi{10.1088/0004-6256/147/4/83}

\bibitem[{{Tothill} {et~al.}(2008){Tothill}, {Gagn{\'e}}, {Stecklum}, \&
  {Kenworthy}}]{tothill2008}
{Tothill}, N.~F.~H., {Gagn{\'e}}, M., {Stecklum}, B., \& {Kenworthy}, M.~A.
  2008, in Handbook of Star Forming Regions, Volume II, ed. B.~{Reipurth},
  Vol.~5, 533

\bibitem[{{Townsley} {et~al.}(2011{\natexlab{a}}){Townsley}, {Broos}, {Chu},
  {Gruendl}, {Oey}, \& {Pittard}}]{townsley2011b}
{Townsley}, L.~K., {Broos}, P.~S., {Chu}, Y.-H., {et~al.} 2011{\natexlab{a}},
  \apjs, 194, 16, \dodoi{10.1088/0067-0049/194/1/16}

\bibitem[{{Townsley} {et~al.}(2011{\natexlab{b}}){Townsley}, {Broos},
  {Corcoran}, {Feigelson}, {Gagn{\'e}}, {Montmerle}, {Oey}, {Smith}, {Garmire},
  {Getman}, {Povich}, {Remage Evans}, {Naz{\'e}}, {Parkin}, {Preibisch},
  {Wang}, {Wolk}, {Chu}, {Cohen}, {Gruendl}, {Hamaguchi}, {King}, {Mac Low},
  {McCaughrean}, {Moffat}, {Oskinova}, {Pittard}, {Stassun}, {ud-Doula},
  {Walborn}, {Waldron}, {Churchwell}, {Nichols}, {Owocki}, \&
  {Schulz}}]{Townsley2011}
{Townsley}, L.~K., {Broos}, P.~S., {Corcoran}, M.~F., {et~al.}
  2011{\natexlab{b}}, \apjs, 194, 1, \dodoi{10.1088/0067-0049/194/1/1}

\bibitem[{{Venuti} {et~al.}(2021){Venuti}, {Cody}, {Rebull}, {Beccari},
  {Irwin}, {Thanvantri}, {Howell}, \& {Barentsen}}]{Venuti2021}
{Venuti}, L., {Cody}, A.~M., {Rebull}, L.~M., {et~al.} 2021, \aj, 162, 101,
  \dodoi{10.3847/1538-3881/ac0536}

\bibitem[{{Venuti} {et~al.}(2014){Venuti}, {Bouvier}, {Flaccomio}, {Alencar},
  {Irwin}, {Stauffer}, {Cody}, {Teixeira}, {Sousa}, {Micela}, {Cuillandre}, \&
  {Peres}}]{Venuti2014}
{Venuti}, L., {Bouvier}, J., {Flaccomio}, E., {et~al.} 2014, \aap, 570, A82,
  \dodoi{10.1051/0004-6361/201423776}

\bibitem[{{Venuti} {et~al.}(2015){Venuti}, {Bouvier}, {Irwin}, {Stauffer},
  {Hillenbrand}, {Rebull}, {Cody}, {Alencar}, {Micela}, {Flaccomio}, \&
  {Peres}}]{venuti2015}
{Venuti}, L., {Bouvier}, J., {Irwin}, J., {et~al.} 2015, \aap, 581, A66,
  \dodoi{10.1051/0004-6361/201526164}

\bibitem[{{Venuti} {et~al.}(2017){Venuti}, {Bouvier}, {Cody}, {Stauffer},
  {Micela}, {Rebull}, {Alencar}, {Sousa}, {Hillenbrand}, \&
  {Flaccomio}}]{venuti2017}
{Venuti}, L., {Bouvier}, J., {Cody}, A.~M., {et~al.} 2017, \aap, 599, A23,
  \dodoi{10.1051/0004-6361/201629537}

\bibitem[{{Vernet} {et~al.}(2011){Vernet}, {Dekker}, {D'Odorico}, {Kaper},
  {Kjaergaard}, {Hammer}, {Randich}, {Zerbi}, {Groot}, {Hjorth}, {Guinouard},
  {Navarro}, {Adolfse}, {Albers}, {Amans}, {Andersen}, {Andersen}, {Binetruy},
  {Bristow}, {Castillo}, {Chemla}, {Christensen}, {Conconi}, {Conzelmann},
  {Dam}, {de Caprio}, {de Ugarte Postigo}, {Delabre}, {di Marcantonio},
  {Downing}, {Elswijk}, {Finger}, {Fischer}, {Flores}, {Fran{\c{c}}ois},
  {Goldoni}, {Guglielmi}, {Haigron}, {Hanenburg}, {Hendriks}, {Horrobin},
  {Horville}, {Jessen}, {Kerber}, {Kern}, {Kiekebusch}, {Kleszcz}, {Klougart},
  {Kragt}, {Larsen}, {Lizon}, {Lucuix}, {Mainieri}, {Manuputy}, {Martayan},
  {Mason}, {Mazzoleni}, {Michaelsen}, {Modigliani}, {Moehler}, {M{\o}ller},
  {Norup S{\o}rensen}, {N{\o}rregaard}, {P{\'e}roux}, {Patat}, {Pena}, {Pragt},
  {Reinero}, {Rigal}, {Riva}, {Roelfsema}, {Royer}, {Sacco}, {Santin},
  {Schoenmaker}, {Spano}, {Sweers}, {Ter Horst}, {Tintori}, {Tromp}, {van
  Dael}, {van der Vliet}, {Venema}, {Vidali}, {Vinther}, {Vola}, {Winters},
  {Wistisen}, {Wulterkens}, \& {Zacchei}}]{vernet2011}
{Vernet}, J., {Dekker}, H., {D'Odorico}, S., {et~al.} 2011, \aap, 536, A105,
  \dodoi{10.1051/0004-6361/201117752}

\bibitem[{{Vrba} {et~al.}(1993){Vrba}, {Chugainov}, {Weaver}, \&
  {Stauffer}}]{vrba1993}
{Vrba}, F.~J., {Chugainov}, P.~F., {Weaver}, W.~B., \& {Stauffer}, J.~S. 1993,
  \aj, 106, 1608, \dodoi{10.1086/116751}

\bibitem[{{Wang} {et~al.}(2011){Wang}, {Feigelson}, {Townsley}, {Broos},
  {Getman}, {Wolk}, {Preibisch}, {Stassun}, {Moffat}, {Garmire}, {King},
  {McCaughrean}, \& {Zinnecker}}]{wang2011}
{Wang}, J., {Feigelson}, E.~D., {Townsley}, L.~K., {et~al.} 2011, \apjs, 194,
  11, \dodoi{10.1088/0067-0049/194/1/11}

\bibitem[{{Yao} {et~al.}(2019){Yao}, {Pepper}, {Gaudi}, {Labadie-Bartz},
  {Beatty}, {Col{\'o}n}, {James}, {Kuhn}, {Lund}, {Rodriguez}, {Siverd},
  {Stassun}, {Stevens}, {Villanueva}, \& {Bayliss}}]{Yao2019}
{Yao}, X., {Pepper}, J., {Gaudi}, B.~S., {et~al.} 2019, \aj, 157, 37,
  \dodoi{10.3847/1538-3881/aaf23c}

\bibitem[{{Zanni} \& {Ferreira}(2009)}]{zanni2009}
{Zanni}, C., \& {Ferreira}, J. 2009, \aap, 508, 1117,
  \dodoi{10.1051/0004-6361/200912879}

\bibitem[{{Zanni} \& {Ferreira}(2013)}]{zanni2013}
---. 2013, \aap, 550, A99, \dodoi{10.1051/0004-6361/201220168}

\end{thebibliography}
\bibliographystyle{aasjournal}

\end{document}